\documentstyle[multicol,epsf,aps,rotate]{revtex}

\begin{document}

\draft

\title{Scaling of the distribution of fluctuations\\
 of financial market indices}

\author{Parameswaran Gopikrishnan$^{1}$, Vasiliki Plerou$^{1,2}$,
        Lu\'{\i}s A.\ Nunes Amaral$^{1}$, \\ Martin Meyer$^{1}$, and
        H. Eugene Stanley$^{1}$}

\address{ $^{1}$ Center for Polymer Studies and Dept. of Physics, Boston
        University, Boston, MA 02215, USA \\ $^{2}$Department of
        Physics, Boston College, Chestnut Hill, MA
        02167, USA }


\maketitle

\begin{abstract}

We study the distribution of fluctuations over a time scale
$\Delta t$ (i.e., the returns) of the S\&P 500 index by analyzing
three distinct databases. Database (i) contains approximately 1
million records sampled at 1~min intervals for the 13-year period
1984--1996, database (ii) contains 8686 daily records for the 35-year
period 1962--1996, and database (iii) contains 852 monthly records for
the 71-year period 1926--1996. We compute the probability
distributions of returns over a time scale $\Delta t$, where $\Delta
t$ varies approximately over a factor of $10^4$---from 1$\,$min up to
more than 1$\,$month. We find that the distributions for $\Delta t
\leq$ 4~days (1560~mins) are consistent with a power-law asymptotic
behavior, characterized by an exponent $\alpha \approx 3$, well
outside the stable L\'evy regime $0 < \alpha < 2$. To test the
robustness of the S\&P result, we perform a parallel analysis on two
other financial market indices.  Database (iv) contains 3560 daily
records of the NIKKEI index for the 14-year period 1984-97, and
database (v) contains 4649 daily records of the Hang-Seng index for
the 18-year period 1980-97. We find estimates of $\alpha$ consistent
with those describing the distribution of S\&P 500 daily-returns. One
possible reason for the scaling of these distributions is the long
persistence of the autocorrelation function of the volatility. For
time scales longer than $(\Delta t)_{\times} \approx 4$~days, our
results are consistent with slow convergence to Gaussian behavior.

\end{abstract}

\begin{multicols}{2}

\section{Introduction and Background}

The analysis of financial data by methods developed for physical
systems has a long
tradition\cite{Bachelier,Pareto1897,Levi37,Mandelbrot63} and has
recently attracted the interest of
physicists\cite{Bouchaud98,ms,Kondor98,palermo,kent,Mantegna95,Bouchaud94,
Sornette96,Ghasghaie96,Arnoedo98,Ausloos,Lux99,Solomon94,Bouchaud98a,
Bouchaud98b,Sornette99,Dietrich,Zhang,Takayasu97}.  Among the reasons
for this interest is the scientific challenge of understanding the
dynamics of a strongly fluctuating complex system with a large number
of interacting elements.  In addition, it is possible that the
experience gained by studying complex physical systems might yield new
results in economics.

Financial markets are complex dynamical systems with many interacting
elements that can be grouped into two categories: (i) the {\it
traders} --- such as individual investors, mutual funds, brokerage
firms, and banks --- and (ii) the {\it assets} --- such as bonds,
stocks, futures, and options. Interactions between these elements lead
to transactions mediated by the stock exchange. The details of each
transaction are recorded for later analysis. The dynamics of a
financial market are difficult to understand not only because of the
complexity of its internal elements but also because of the many
intractable external factors acting on it, which may even differ from
market to market. Remarkably, the statistical properties of certain
observables appear to be similar for quite different
markets\cite{Pagan96,Mantegna91,Skjeltrop}, consistent with the
possibility that there may exist ``universal'' results.

The most challenging difficulty in the study of a financial market is
that the nature of the interactions between the different elements
comprising the system is unknown, as is the way in which external
factors affect it.  Therefore, as a starting point, one may resort to
empirical studies to help uncover the regularities or ``empirical
laws'' that may govern financial markets.

\begin{figure}  
\narrowtext 
\centerline{
\epsfysize=0.8\columnwidth{\rotate[r]{\epsfbox{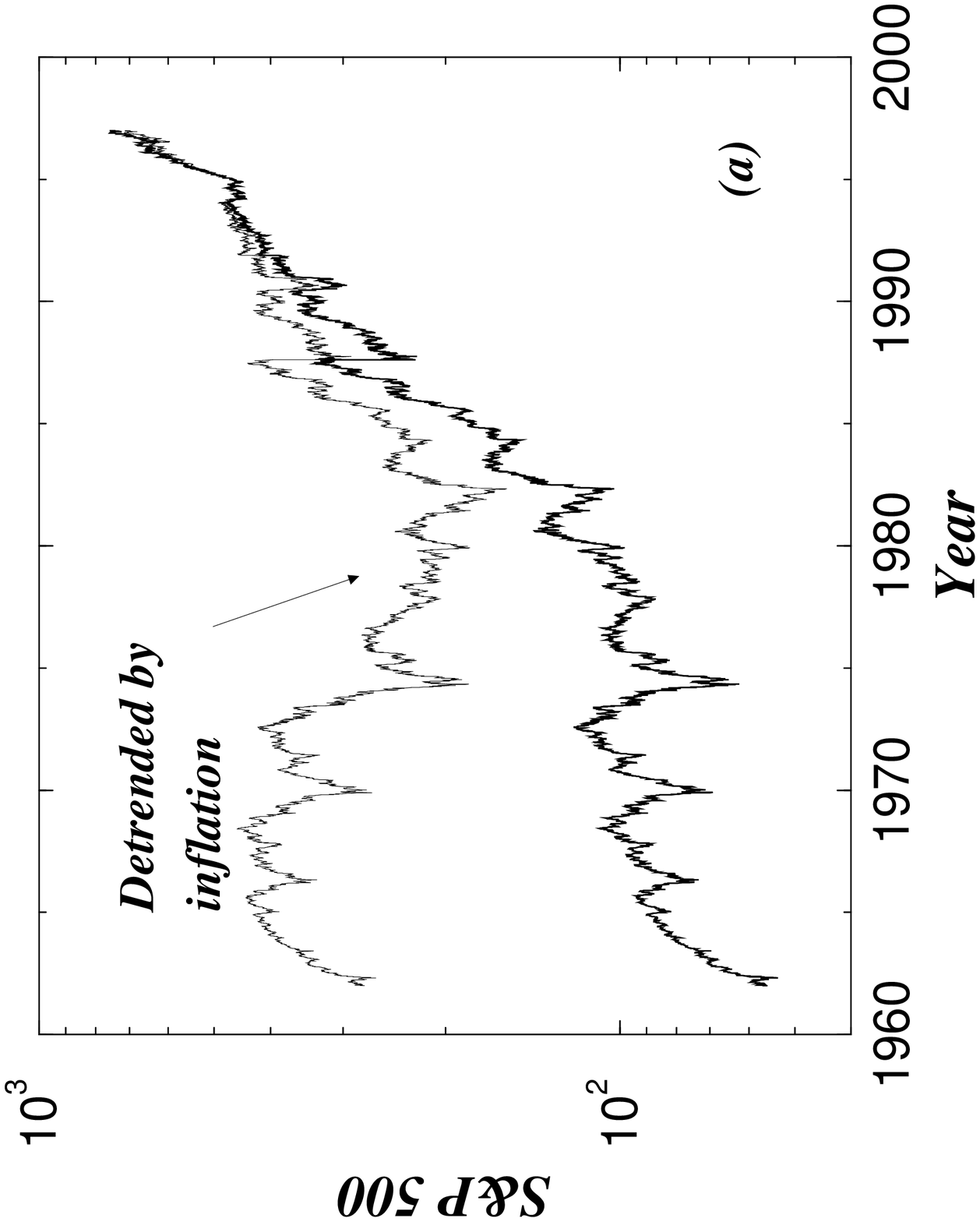}}} 
}
\centerline{
\epsfysize=0.8\columnwidth{\rotate[r]{\epsfbox{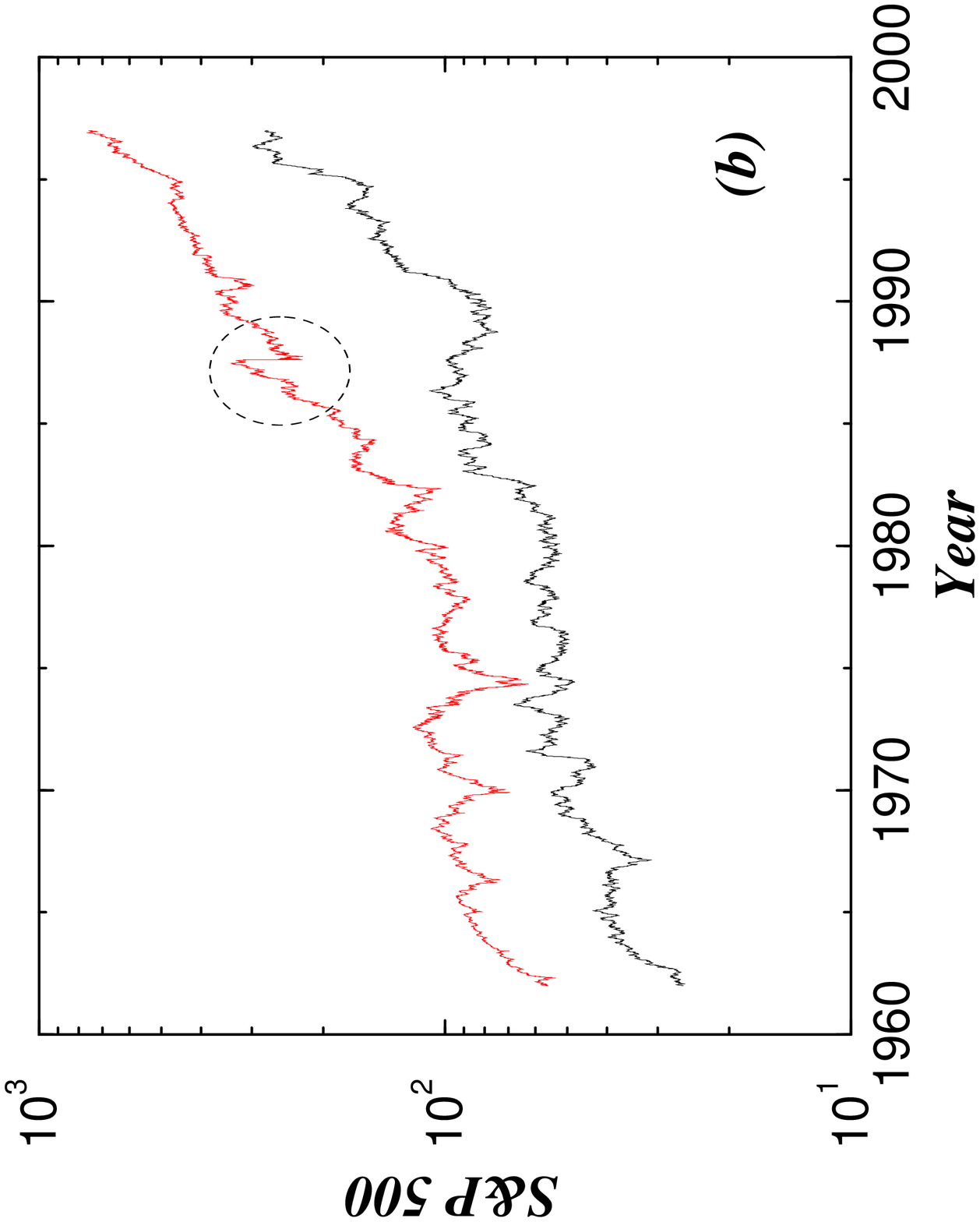}}} 
}
\caption{The S\&P 500 index is the sum of the market capitalizations
of 500 companies. In {\it (a)}, we display both the value of the S\&P
500 index (bottom line) and the index detrended by inflation to 1994
US dollars (top line). The sharp jump seen in 1987 is the market crash
of October 19. {\it (b)} Comparison of the time evolution of the S\&P
500 for the 35-year period 1962--96 (top line) and a biased Gaussian
random walk (bottom line). The random walk has the same bias as the
S\&P 500 ---approximately 7\% per year for the period considered.}
\label{sp500_brownian}
\end{figure}

The interactions between the different elements comprising financial
markets generate many observables such as the transaction price, the
share volume traded, the trading frequency, and the values of market
indices [Fig.~\ref{sp500_brownian}]. A number of studies investigated
the time series of returns on varying time scales $\Delta t$ in order
to probe the nature of the stochastic process underlying
it~\cite{Mantegna95,Bouchaud94,Sornette96,Ghasghaie96,Arnoedo98,Ausloos,
Dacarogna93,Lux98}.
For a time series $S(t)$ of prices or market index values, the return
$G(t) \equiv G_{\Delta t}(t)$ over a time scale $\Delta t$ is defined
as the forward change in the logarithm of $S(t)$\cite{frc},
\begin{equation}
G_{\Delta t}(t)\equiv \ln S(t+\Delta t) - \ln S(t)\,. 
\label{return}
\end{equation}
For small changes in $S(t)$, the return $ G_{\Delta t}(t)$ is
approximately the forward relative change,
\begin{equation}
G_{\Delta t} (t) \approx {S(t+\Delta t)-S(t)\over S(t)}.
\label{relChange}
\end{equation}

\begin{figure}[hbt]
\narrowtext
\centerline{
\epsfysize=0.9\columnwidth{\rotate[r]{\epsfbox{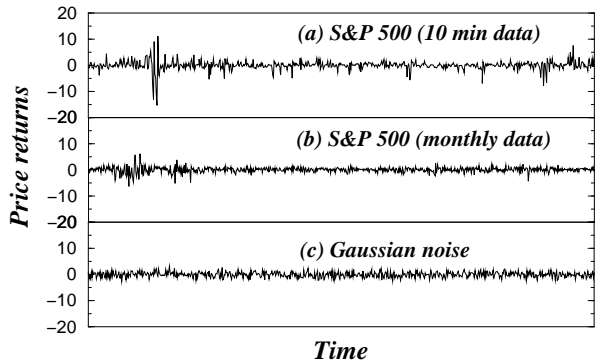}}}
}
\caption{Sequence of {\it (a)} 10$\,$min returns, from database (i),
and {\it (b)} 1$\,$month returns, from database (iii), for the S\&P
500, normalized to unit variance.  {\it (c)} Sequence of {\it
i.i.d.\/} Gaussian random variables with unit variance, which was
proposed by Bachelier as a model for stock
returns\protect\cite{Bachelier}. For all 3 panels, there are 850
events ---i.e., in panel (a) 850 minutes and in panel (b) 850
months. Note that, in contrast to (a) and (b), there are no
``extreme'' events in (c).}
\label{gt_brownian}   
\end{figure}

In 1900, Bachelier proposed the first model for the stochastic process
of returns---an uncorrelated random walk with independent, identically
Gaussian distributed ({\it i.i.d\/}) random variables\cite{Bachelier}.
This model is natural if one considers the return over a time scale
$\Delta t$ to be the result of many independent ``shocks'', which then
lead by the central limit theorem to a Gaussian distribution of
returns~\cite{Bachelier}.  However, empirical studies
\cite{Mandelbrot63,Mantegna95,Bouchaud94,Sornette96,Ghasghaie96} show
that the distribution of returns\cite{Nonstationarity} has pronounced
tails in striking contrast to that of a Gaussian.  To illustrate this
fact, we show in Fig.~\ref{gt_brownian} the 10$\,$min returns of the
S\&P 500 market index\cite{sp_comm} for 1986-1987 and contrast it with
a sequence of {\it i.i.d.} Gaussian random variables. Both are
normalized to have unit variance. Clearly, large events are very
frequent in the data, a fact largely underestimated by a Gaussian
process. Despite this empirical fact, the Gaussian assumption for the
distribution of returns is widely used in theoretical finance because
of the simplifications it provides in analytical calculation; indeed,
it is one of the assumptions used in the classic Black-Scholes option
pricing formula\cite{bs}.

In his pioneering analysis of cotton prices, Mandelbrot observed that
in addition to being non-Gaussian, the process of returns shows
another interesting property: ``time scaling'' --- that is, the
distributions of returns for various choices of $\Delta t$, ranging
from 1~day up to 1~month have similar functional
forms\cite{Mandelbrot63}. Motivated by (i) pronounced tails, and (ii)
a stable functional form for different time scales,
Mandelbrot\cite{Mandelbrot63} proposed that the distribution of
returns is consistent with a L\'evy stable
distribution\cite{Pareto1897,Levi37} --- that is, the returns can be
modeled as a L\'evy stable process. L\'evy stable distributions arise
from the generalization of the central limit theorem to random
variables which do not have a finite second moment [see Appendix A].

Conclusive results on the distribution of returns are difficult to
obtain, and require a large amount of data to study the rare events
that give rise to the tails. More recently, the availability of high
frequency data on financial market indices, and the advent of improved
computing capabilities, has facilitated the probing of the asymptotic
behavior of the distribution.  For these reasons, recent empirical
studies\cite{Mantegna95,Bouchaud94,Sornette96,Ghasghaie96} of the S\&P
500 index\cite{sp_comm} analyze typically $10^6$--$10^7$ data points,
in contrast to approximately 2000 data points analyzed in the classic
work of Mandelbrot\cite{Mandelbrot63}. Reference~\cite{Mantegna95}
reports that the central part of the distribution of S\&P 500
returns appears to be well fit by a L\'evy distribution, but the
asymptotic behavior of the distribution of returns shows faster decay
than predicted by a L\'evy distribution. Hence, Ref.~\cite{Mantegna95}
proposed a truncated L\'evy distribution---a L\'evy distribution in
the central part followed by an approximately exponential
truncation---as a model for the distribution of returns. The
exponential truncation ensures the existence of a finite second
moment, and hence the truncated L\'evy distribution is not a stable
distribution~\cite{Mantegna94,Koponen}. The truncated L\'evy process
with {\it i.i.d.\/} random variables has slow convergence to Gaussian
behavior due to the L\'evy distribution in the center, which could
explain the observed time scaling for a considerable range of time
scales\cite{Mantegna95}.

\begin{figure}[hbt]
\narrowtext 
\centerline{
\epsfysize=0.8\columnwidth{\rotate[r]{\epsfbox{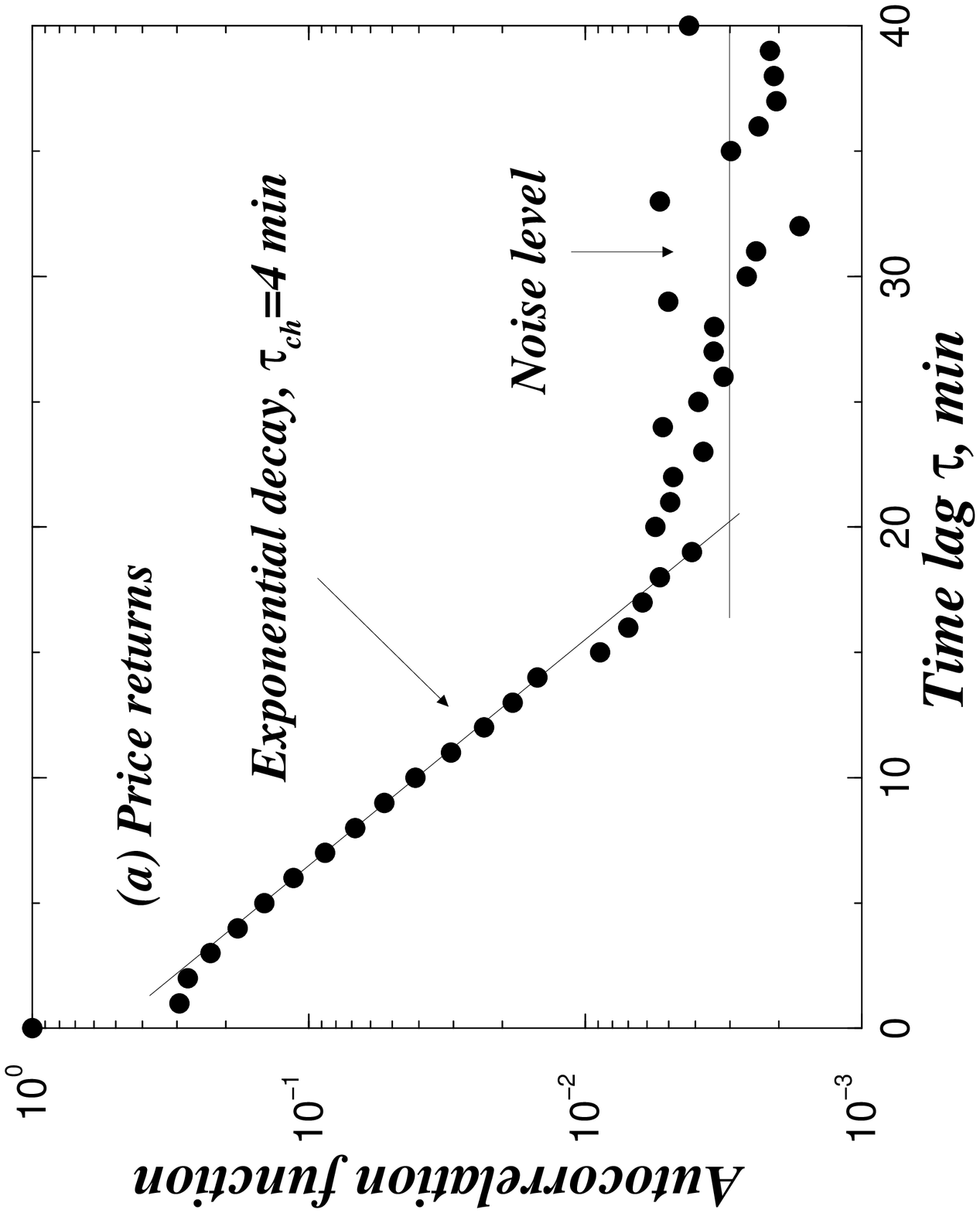}}}
}
\centerline{
\epsfysize=0.8\columnwidth{\rotate[r]{\epsfbox{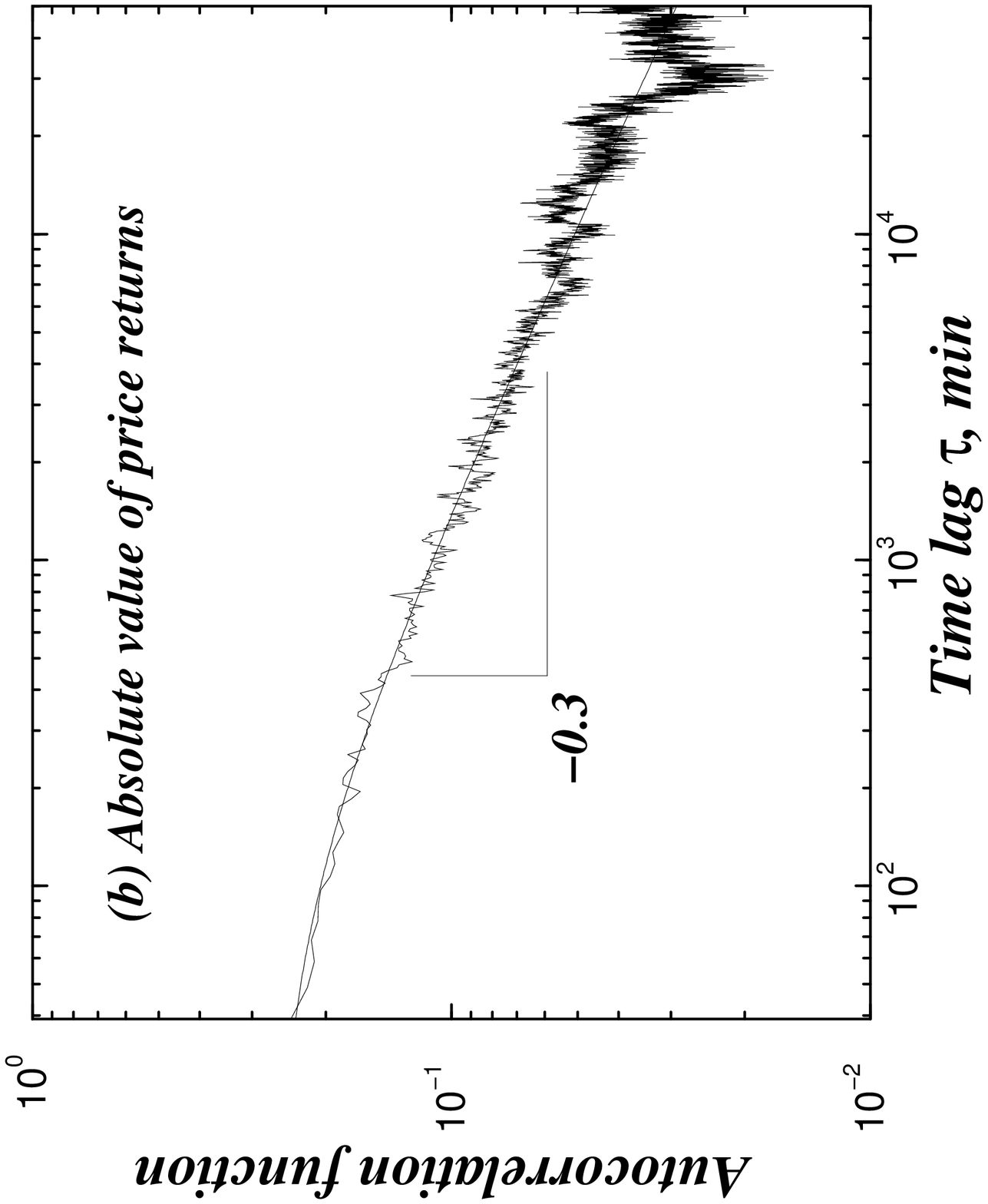}}}
}
\centerline{
\epsfysize=0.8\columnwidth{\rotate[r]{\epsfbox{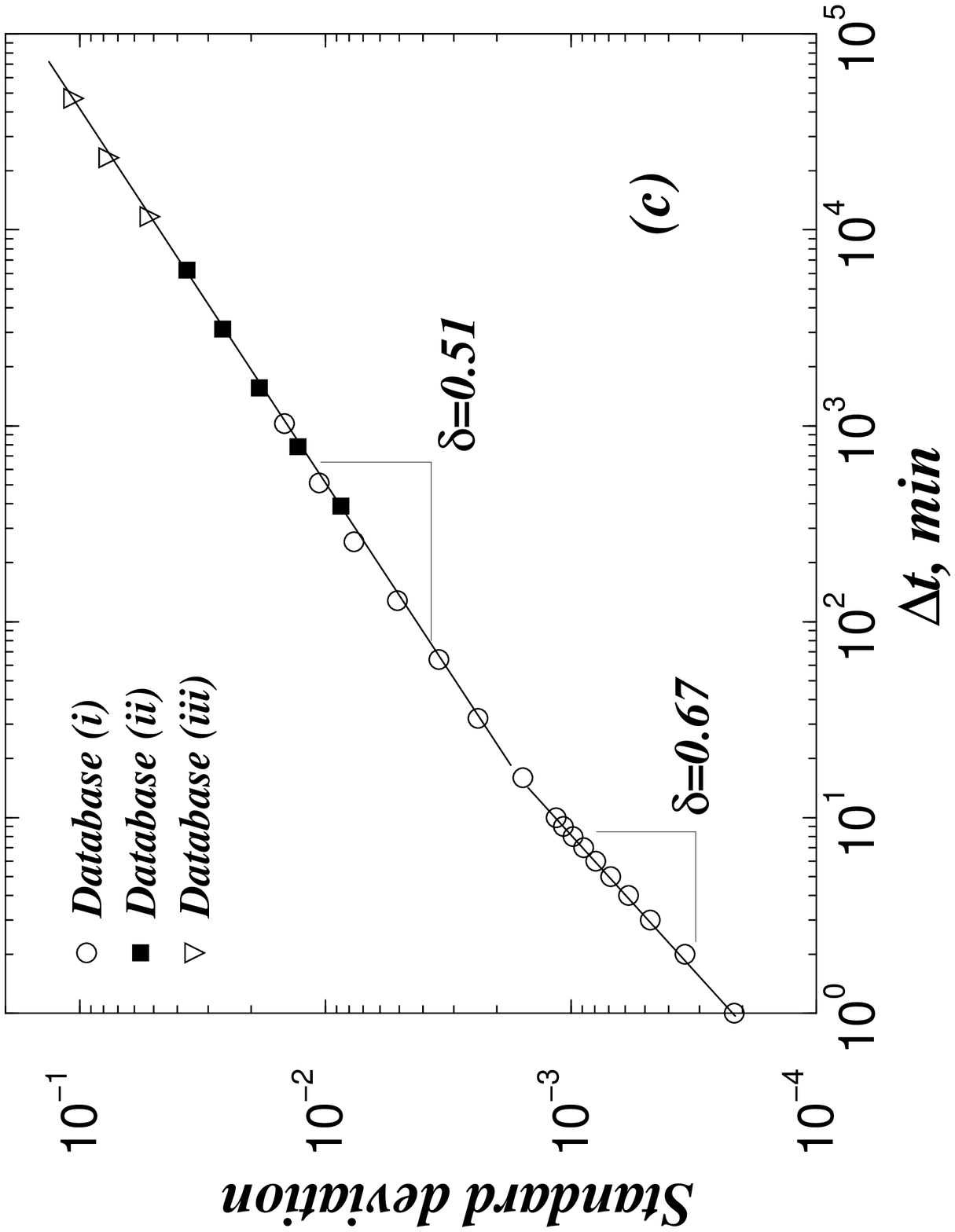}}}
}
\caption{{\it (a)} Semilog plot of the autocorrelation function for
the S\&P 500 returns $G_{\Delta t}(t)$ sampled at a $\Delta t =1\,$min
time scale, $C_{\Delta t}(\tau) \equiv [\langle G_{\Delta
t}(t)\,G_{\Delta t}(t+\tau) \rangle - \langle G_{\Delta
t}(t)\rangle^2] / [ \langle G_{\Delta t}(t)^2 \rangle -\langle
G_{\Delta t}(t) \rangle^2 ]$.  The straight line corresponds to an
exponential decay with a characteristic decay time $\tau_{ch} =
4$~min. Note that after 20$\,$min the correlations are at the noise
level. {\it (b)} Loglog plot of the autocorrelation function of the
absolute returns. The solid line is a power-law regression fit over
the entire range, which gives an estimate of the power-law exponent,
$\eta=0.29\pm 0.05$. Better estimates of this exponent can be obtained
from the power spectrum or from other more sophisticated methods. It
has been recently reported using such methods that the autocorrelation
function of the absolute value of the returns shows {\it two\/}
power-law regimes with a crossover at approximately
1.5~days\protect\cite{Yanhui97}. (c) Loglog plot of the time averaged
volatility $v \equiv v(\Delta t)$ as a function of the time scale
$\Delta t$ of the returns obtained from databases (i--iii). For
$\Delta t \leq 20$~min, we observe a slope $\delta=0.67 \pm 0.03$, due
to the exponentially-damped time correlations. For $\Delta t \ge
20$~min, we observe $\delta = 0.51\pm 0.06$, indicating the absence of
significant correlations.  }
\label{acorsp}
\end{figure}

In addition to the probability distribution, a complementary aspect
for the characterization of any stochastic process is the
quantification of correlations.  Studies of the autocorrelation
function of returns show exponential decay with characteristic decay
times $\tau_{ch}$ of only 4$\,$min\cite{Dacarogna93,FXcorr,Ding93}.
As is clear from Fig.~\ref{acorsp}(a), for time scales beyond 20
$\,$min the correlation function is at the level of noise, in
agreement with the {\it efficient market hypothesis\/} which states
that is not possible to predict future stock prices from their
previous values\cite{Fama70}. If price-correlations were not
short-range, one could devise a way to make money from the market
indefinitely.

It is important to note that lack of linear correlation does not imply
an {\it i.i.d.\/} process for the returns, since there may exist
higher-order correlations [Fig~\ref{acorsp}(b)]. Indeed, the amplitude
of the returns, referred to in economics as the {\it
volatility\/}\cite{volatility}, shows long-range time correlations
that persist up to several
months\cite{Dacarogna93,Ding93,Fama70,volatility,Granger96,
Yanhui97,Yanhui98,pierre,Cont97,Pasquini,Beran}, and are characterized
by an asymptotic power-law decay.

\section{Motivation}

A recent preliminary study reported that the distributions of 5$\,$min
returns for 1000 individual stocks and the S\&P 500 index decay as a
power-law with an exponent well outside the stable L\'evy
regime\cite{Gopi98}. Consistent results were found by studies both on
stock markets~\cite{Pagan96} and on foreign exchange
markets\cite{Olsen}.  These results raise two important questions:

First, the distribution of returns has a finite second moment, thus,
we would expect it to converge to a Gaussian because of the central
limit theorem.  On the other hand, preliminary studies suggest the
distributions of returns retain their power-law functional form for
long time scales.  So, we can ask which of these two scenarios is
correct?  We find that the distributions of returns retain their
functional form for time scales up to approximately 4~days, after
which we find results consistent with a slow convergence to Gaussian
behavior.

Second, power-law distributions are not stable distributions, but the
distribution of returns retains its functional form for a range of
time scales.  It is then natural to ask how can this {\it scaling
behavior\/} possibly arise?  One possible explanation is the
recently-proposed exponentially-truncated L\'evy
distribution\cite{Mantegna95,Mantegna94,Koponen}. However, the
truncated L\'evy process is constructed out of {\it i.i.d.} random
variables and hence is not consistent with the empirically-observed
long persistence in the autocorrelation function of the volatility of
returns~\cite{Dacarogna93,Ding93,Fama70,volatility,Granger96,Yanhui97,
Yanhui98,pierre,Cont97,Pasquini}. Moreover, our data support the
possibility that the asymptotic nature of the distribution is a
power-law with an exponent outside the L\'evy regime. Also, we will
argue that the scaling behavior observed in the distribution of
returns may be connected to the slow decay of the volatility
correlations.

The organization of the paper is as follows. Section III describes the
data analyzed. Sections IV and V study the distribution of returns
of the S\&P 500 index on time scales $\Delta t \leq 1$~day and $\Delta
t > 1$~day, respectively. Section VI discusses how time correlations in
volatility are related to the time scaling of the distributions, and
Sect.\ VII presents concluding remarks.

\section{The Data analyzed}

First, we analyze the S\&P 500 index, which comprises 500 companies
chosen for market size, liquidity, and industry group representation
in the US.  The S\&P 500 is a market-value weighted index (stock price
times number of shares outstanding), with each stock's weight
proportional to its market value.  The S\&P 500 index is one of the
most widely used benchmarks of U.S. equity performance. In our study,
we first analyze database (i) which contains ``high-frequency'' data
that covers the 13 years period 1984--1996, with a recording frequency
of less than 1~min. The total number of records in this database
exceeds $4.5 \times 10^6$. To investigate longer time scales, we study
two other databases.  Database (ii) contains daily records of the S\&P
500 index for the 35-year period 1962--1996, and database (iii)
contains monthly records for the 71-year period 1926--1996.

In order to test if our results are limited to the S\&P 500 index, we
perform a parallel analysis on two other market indices.  Database
(iv) contains 3560 daily records of the NIKKEI index of the Tokyo
stock exchange for the 14-year period 1984--1997, and database (v)
contains 4649 daily records of the Hang-Seng index of the Hong Kong
stock exchange for the 18-year period 1980--1997.

\section{The distribution of returns for $\Delta t \leq 1$~day}

\subsection{The distribution of returns for $\Delta t = 1$~min}
 
First, we analyze the values of the S\&P500 index from the
high-frequency data for the 13-year period 1984--1996, which extends
the database studied in Ref.~\cite{Mantegna95} by an additional 7
years. The data are typically recorded at 15~second intervals. We
first sample the data at 1~min intervals and generate a time series
$S(t)$ with approximately 1.2~million data points.  From the time
series $S(t)$, we compute the return $G \equiv G_{\Delta t}(t)$ which
is the relative change in the index, defined in Eq.~(\ref{return}).

\begin{figure}
\centerline{  
\epsfysize=0.8\columnwidth{\rotate[r]{\epsfbox{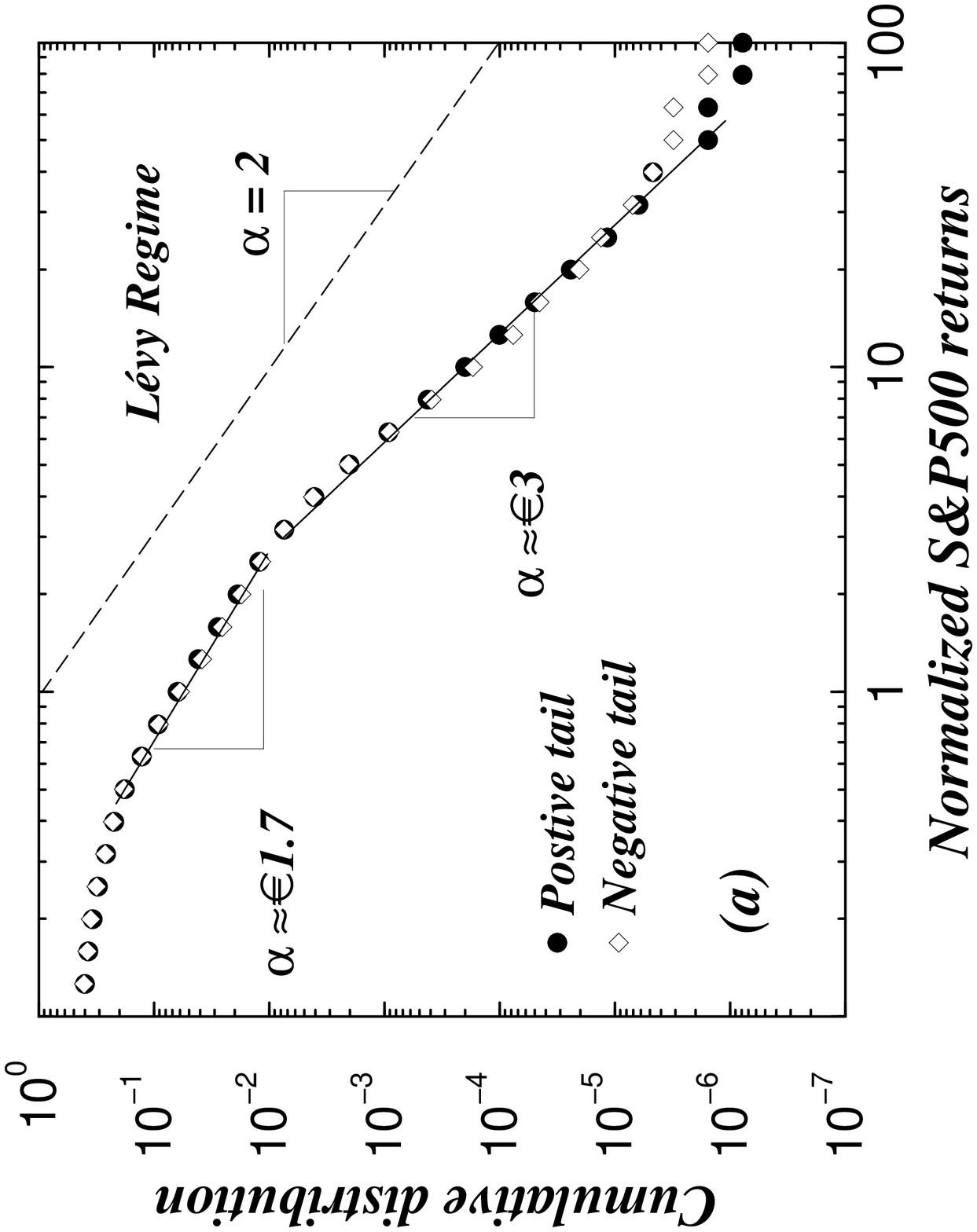}}}
}
\centerline{
\epsfysize=0.8\columnwidth{\rotate[r]{\epsfbox{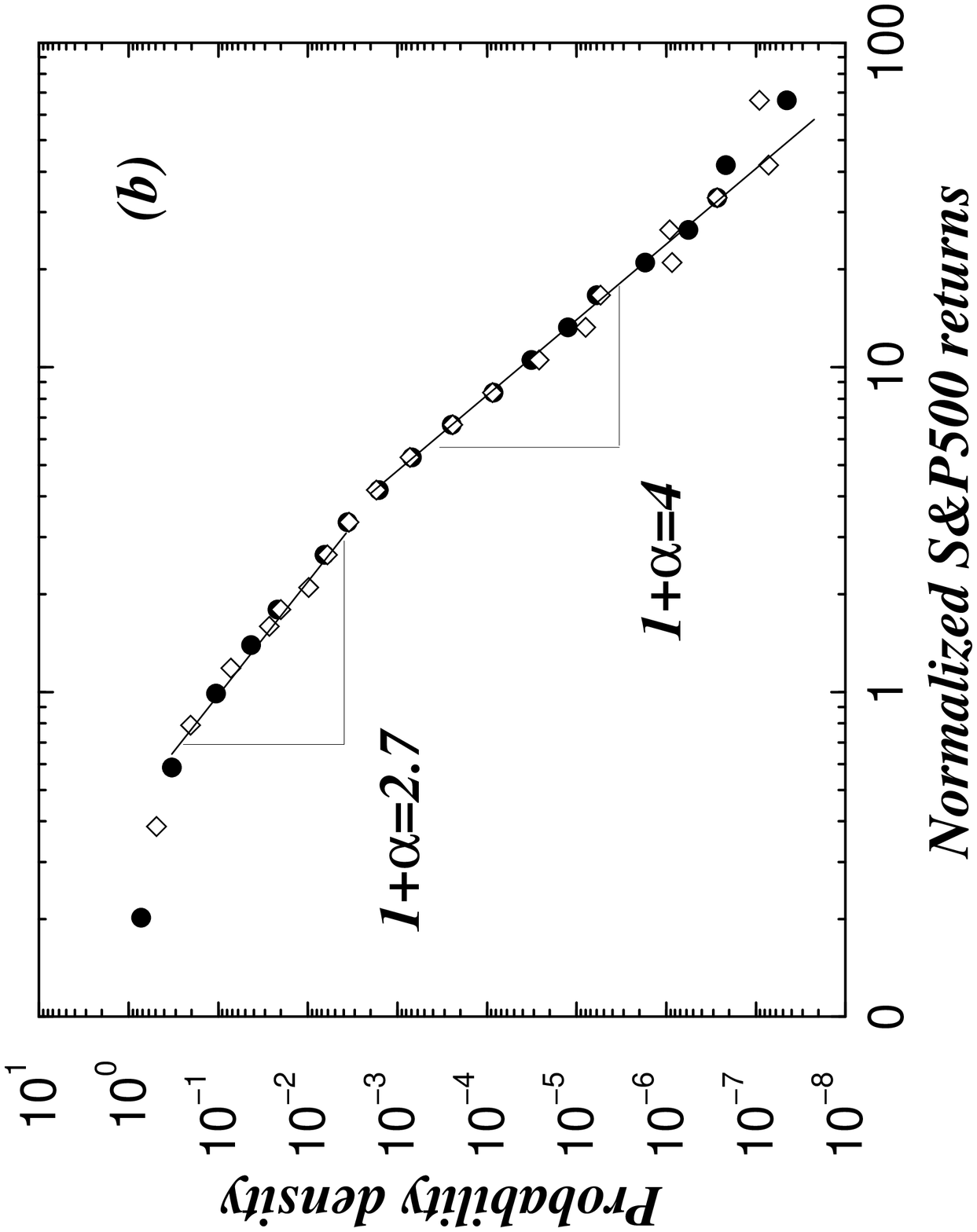}}}
}
\centerline{  
\epsfysize=0.8\columnwidth{\rotate[r]{\epsfbox{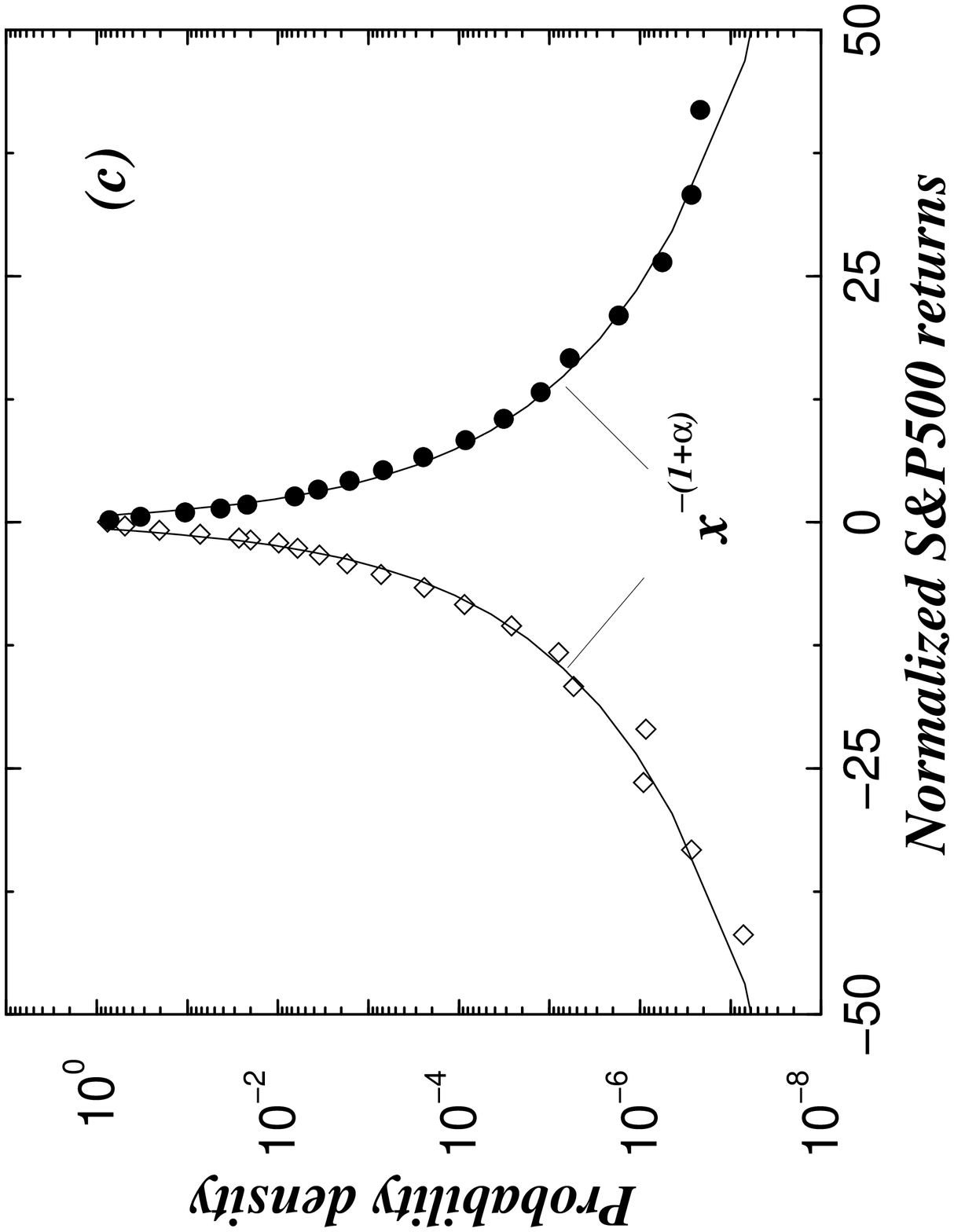}}}  
}
\caption{{\it (a)} Loglog plot of the cumulative distribution of the
normalized 1~min returns for the S\&P 500 index. Power-law regression
fits in the region $3 \leq g \leq 50$ yield $\alpha=2.95\pm 0.07$
(positive tail), and $\alpha=2.75\pm 0.13$ (negative tail). For the
region $0.5 \leq g \leq 3$, regression fits give $\alpha = 1.6 \pm
0.1$ (positive tail), and $\alpha = 1.7 \pm 0.1$ (negative tail).
{\it(b)} Loglog and {\it (c)} linearlog plots of the probability
density function for the normalized S\&P500 returns. The solid lines
are power-law fits with exponents $1+ \alpha \approx 4$. Power-law
regression fits in the region $3\leq g \leq 50$ yield estimates
$\alpha = 3.01 \pm 0.11$ (positive tail), and $\alpha = 3.02 \pm 0.08$
(negative tail). }
\label{sp500_hist}
\end{figure}

In order to compare the behavior of the distribution for different
time scales $\Delta t$, we define a normalized return $g \equiv
g_{\Delta t}(t)$
\begin{equation}
g\equiv {G - \langle G \rangle_T \over v}\,.
\label{gsp.def}
\end{equation}
Here, the time averaged volatility $v \equiv v(\Delta t) $ is defined
through $v^2 \equiv \langle {G}^2 \rangle_T - {\langle G \rangle_T}^2$
and $\langle\dots\rangle_T$ denotes an average over the entire length
of the time series. Figure~\ref{sp500_hist}(a) shows the cumulative
distribution of returns for $\Delta t = 1$~min. For both positive and
negative tails, we find a power-law asymptotic behavior
\begin{equation}
P(g>x) \sim {1\over x^{\alpha}}\,,
\label{def_alpha}
\end{equation} 
similar to what was found for individual stocks\cite{Gopi98}.  For the
region $3 \leq g \leq 50$, regression fits yield
\begin{equation}
\alpha = \cases{ 3.05 \pm 0.04 & (positive tail) \cr
2.94 \pm 0.08 & (negative tail)},
\end{equation}
well outside the L\'evy stable range, $0 \leq \alpha < 2$ . Consistent
values for $\alpha$ are also obtained from the density function. For a
more accurate estimation of the asymptotic behavior, we use the
modified Hill estimator [Fig.~\ref{sp500_hill}(a,b)]. We obtain
estimates for the asymptotic slope in the region $3 \leq g \leq 50$ :
\begin{equation}
\alpha = \cases{ 2.93 \pm 0.11 & (positive tail) \cr
3.02 \pm 0.15 & (negative tail)}.
\end{equation}

For the region $g \leq 3$, regression fits yield smaller estimates of
$\alpha$, consistent with the possibility of a L\'evy distribution in
the central region. The values of $\alpha$ obtained in this range are
quite sensitive to the bounds of the region used for fitting. Our
estimates range from $\alpha \approx 1.35$ up to $\alpha \approx 1.8$
for different fitting regions in the interval $0.1 \leq g \leq 6$. For
example, in the region $0.5 \leq g \leq 3$, we obtain
\begin{equation}
\alpha \approx \cases{ 1.6 & (positive tail) \cr
1.7 & (negative tail)},
\end{equation}
which are consistent with the result $\alpha \approx 1.4$ found for
small values of $g$ in Ref.~\protect\cite{Mantegna95}.  Note that in
Ref.~\protect\cite{Mantegna95} the estimates of $\alpha$ were
calculated using the scaling form of the return probability to the
origin $P(0)$. It is possible that for the financial data analyzed
here, $P(0)$ is not the optimal statistic, because of the discreteness
of the individual-company distributions that comprise
it\cite{discrete}. It is also possible that our values of $\alpha$ for
small values of $g$ could be due to the discreteness in the returns of
the individual companies comprising the S\&P 500.

\subsection{Scaling of the distribution of returns for  
$\Delta t$ up to 1~day}

\begin{figure} 
\centerline{ 
\epsfysize=0.8\columnwidth{\rotate[r]{\epsfbox{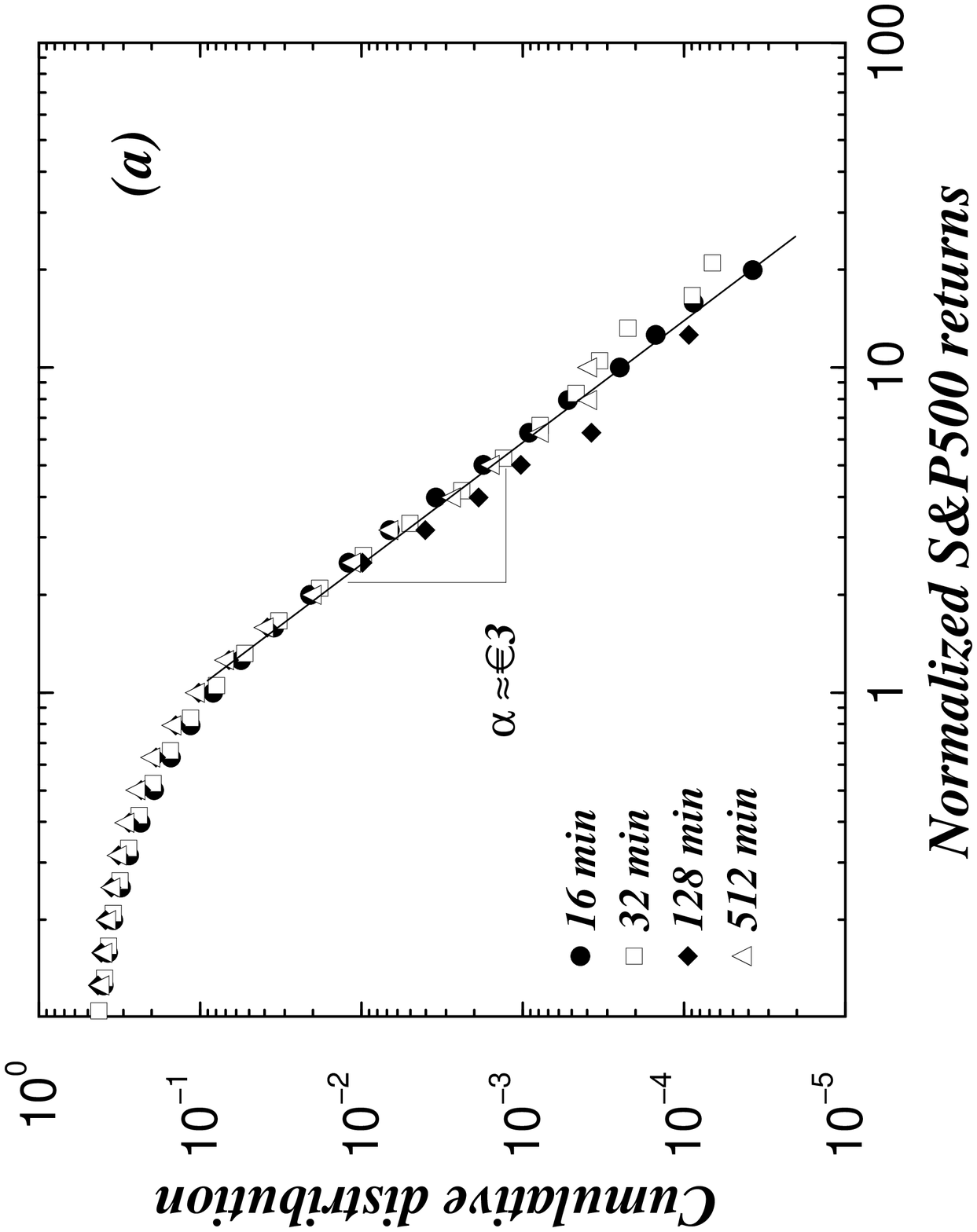}}}
} 
\centerline{
\epsfysize=0.8\columnwidth{\rotate[r]{\epsfbox{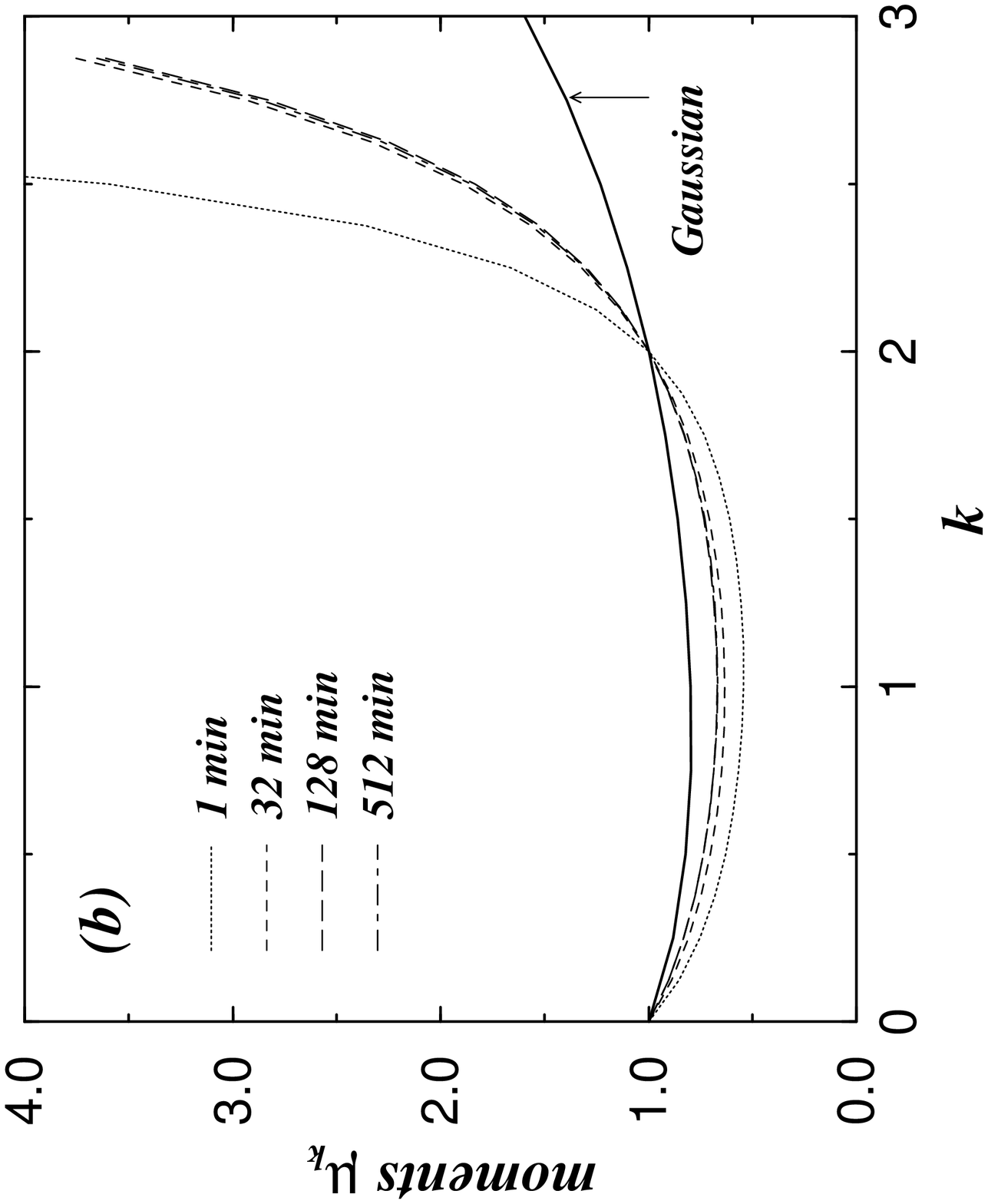}}}  
} 
\caption{ {\it(a)} Loglog plot of the cumulative distribution of
normalized returns of the positive tails for $\Delta t = 16, 32, 128,
512$~mins. Power-law regression fits yield estimates of the asymptotic
power-law exponent $\alpha=2.69 \pm 0.04$, $\alpha =2.53\pm 0.06$,
$\alpha= 2.83 \pm 0.18$ and $\alpha=3.39 \pm 0.03$ for $\Delta t = 16,
32, 128$ and 512~mins, respectively. {\it(b)} The moments of the
distribution for $\Delta t = 1, 32, 128$ and 512~min. The change in
the behavior of the moments from the 1~min scale is probably the
effect of the gradual disappearance of the L\'evy slope for small
values of $g$. For $\Delta t > 30$~min there is no region with slopes
in the L\'evy range, and we observe good agreement between all time
scales.}
\label{sp500_scl} 
\end{figure}

Next, we study the distribution of normalized returns for longer time
scales.  Figure~\ref{sp500_scl}(a) shows the cumulative distribution
of normalized S\&P 500 returns for time scales up to 512~min
(approximately 1.5~days). The distribution appears to retain its
power-law functional form for these time scales.  We verify this
scaling behavior by analyzing the moments of the distribution of
normalized returns $g$,
\begin{equation}
\mu_k \equiv \langle\, \vert g \vert^k \, \rangle_T\,,
\label{moments}
\end{equation}
where $\langle \dots \rangle_T$ denotes an average over all the
normalized returns for all the bins. Since $\alpha \approx 3$, we
expect $\mu_k$ to diverge for $k \geq 3$, and hence we compute $\mu_k$
for $k< 3$.

Figure~\ref{sp500_scl}(b) shows the moments of the normalized returns
$g$ for different time scales from 5~min up to 1~day. The moments do
not vary significantly for the above time scales, confirming the
apparent scaling behavior of the distribution observed in
Fig.~\ref{sp500_scl}(a).

\section{The distribution of returns for $\Delta t \geq 1$~day}

\subsection{The S\&P 500 index}

\begin{figure} 
\centerline{ 
\epsfysize=0.8\columnwidth{\rotate[r]{\epsfbox{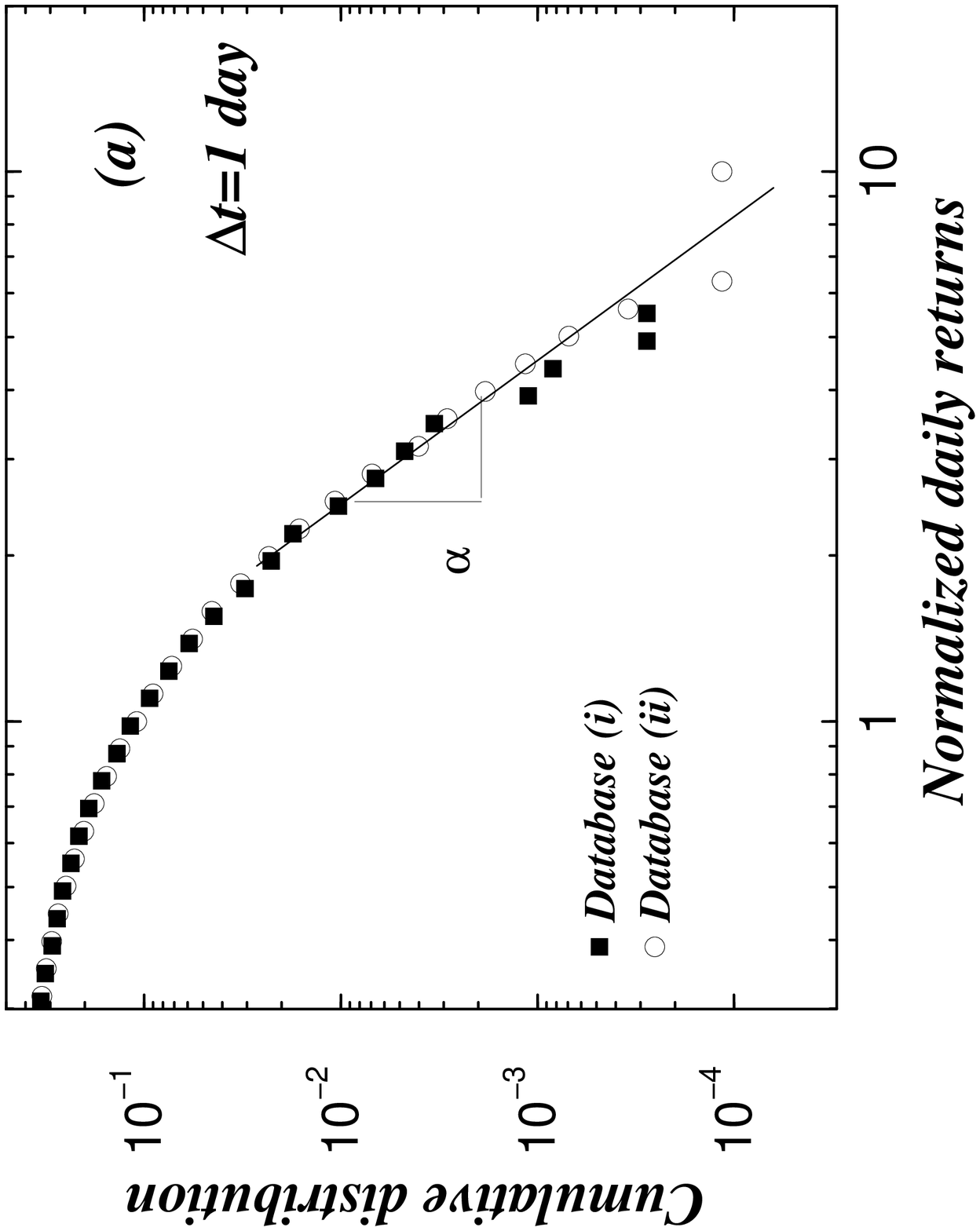}}} 
}
\centerline{
\epsfysize=0.8\columnwidth{\rotate[r]{\epsfbox{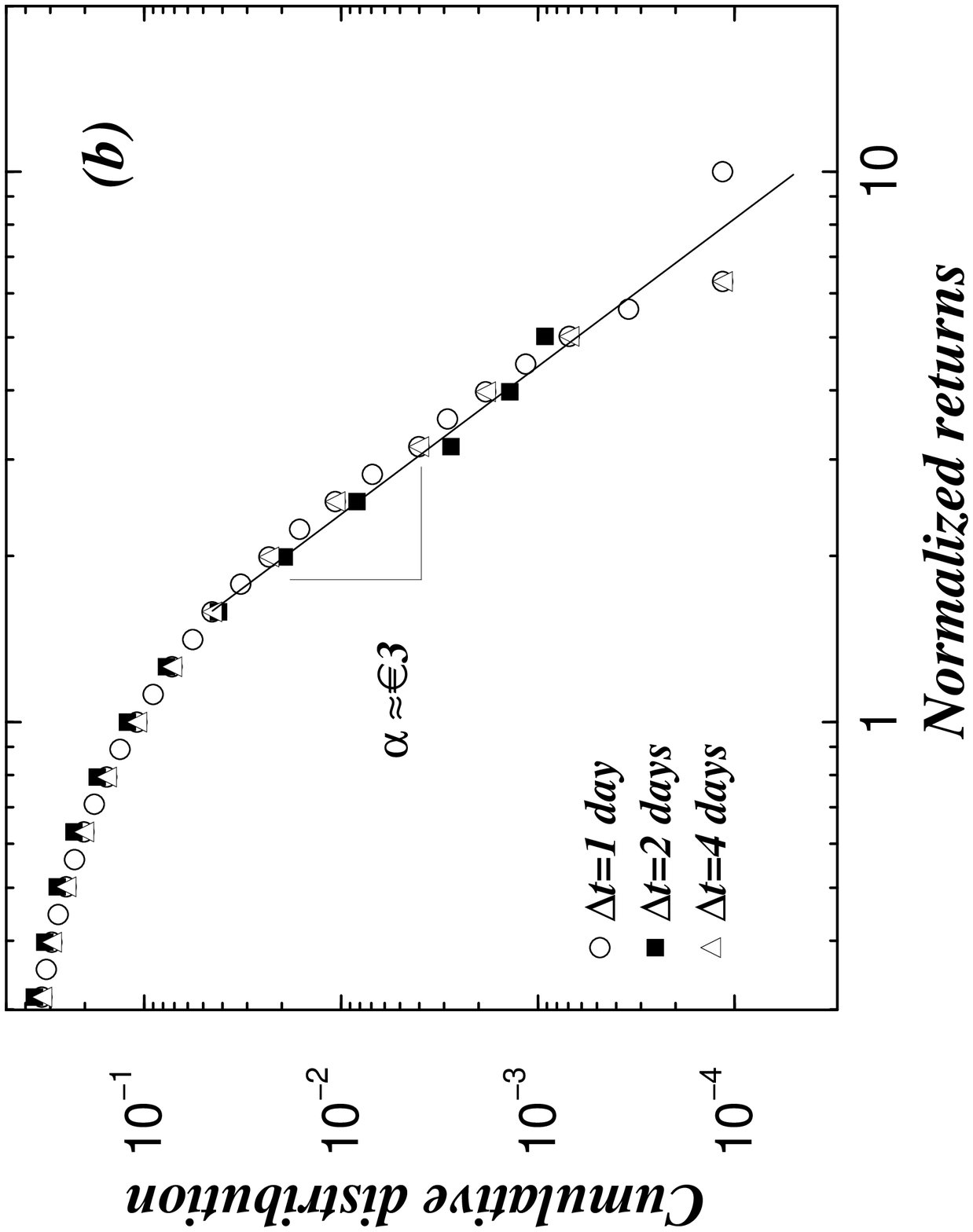}}}  
}
\centerline{ 
\epsfysize=0.8\columnwidth{\rotate[r]{\epsfbox{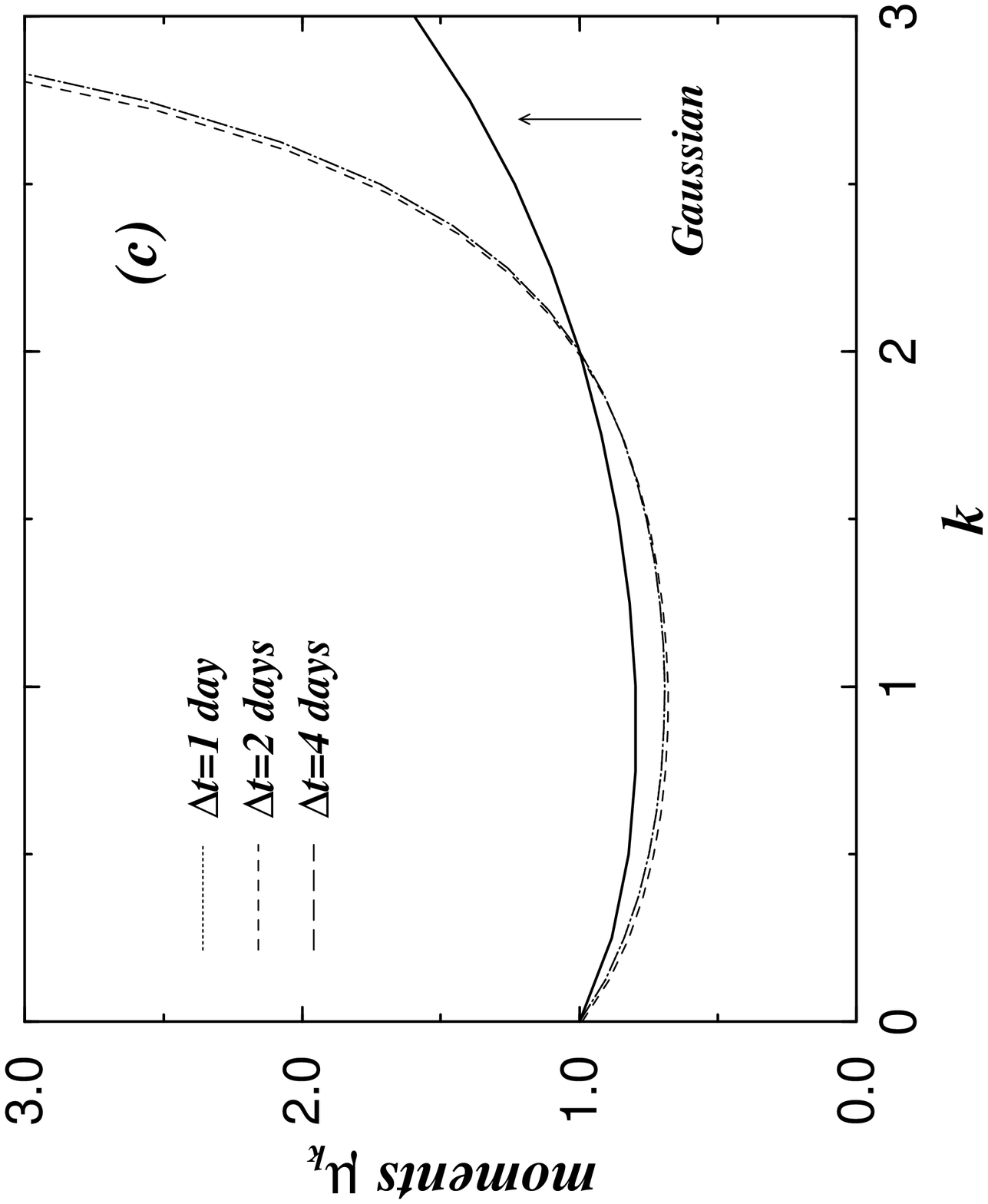}}}
}
\caption{{\it(a)} Cumulative distribution of the normalized S\&P 500
returns from two different databases: Database (i) which contains
1$\,$min records for 13 years, and database (ii) which contains daily
records for 35 years. Power-law regression fits in the region $g \ge
1$ lead to the estimates $\alpha = 3.75 \pm 0.30$ for database (i),
and $\alpha = 3.66 \pm 0.11$ for database (ii).  {\it (b)} The
cumulative distribution from database (ii) for $\Delta t = 1, 2$ and
4~days. The apparent scaling behavior of these distributions is
confirmed by the estimates $\alpha = 3.75 \pm 0.41$ ($\Delta t =
2$~days) and $\alpha = 3.77 \pm 0.29$ ($\Delta t = 4$~days). {\it(c)}
The behavior of the moments for these time scales is in agreement with
the apparent scaling behavior. }
\label{sp500_1d} 
\end{figure}

For time scales beyond 1~day, we use database (ii) which contains
daily-sampled records of the S\&P 500 index for the 35-year period
1962--1996. Figure~\ref{sp500_1d}(a) shows the agreement between
distributions of normalized S\&P 500 daily-returns from database (i),
which contains 1$\,$min sampled data, and database (ii), which
contains daily-sampled data.  Regression fits for the region $1 \leq g
\leq 10$ give estimates of $\alpha \approx 3$.
Figure~\ref{sp500_1d}(b) shows the scaling behavior of the
distribution for $\Delta t =$ 1, 2, and 4~days. For these choices of
$\Delta t$, the scaling behavior is also visible for the moments
[Fig.~\ref{sp500_1d}(c)].

\begin{figure}
\centerline{
\epsfysize=0.8\columnwidth{\rotate[r]{\epsfbox{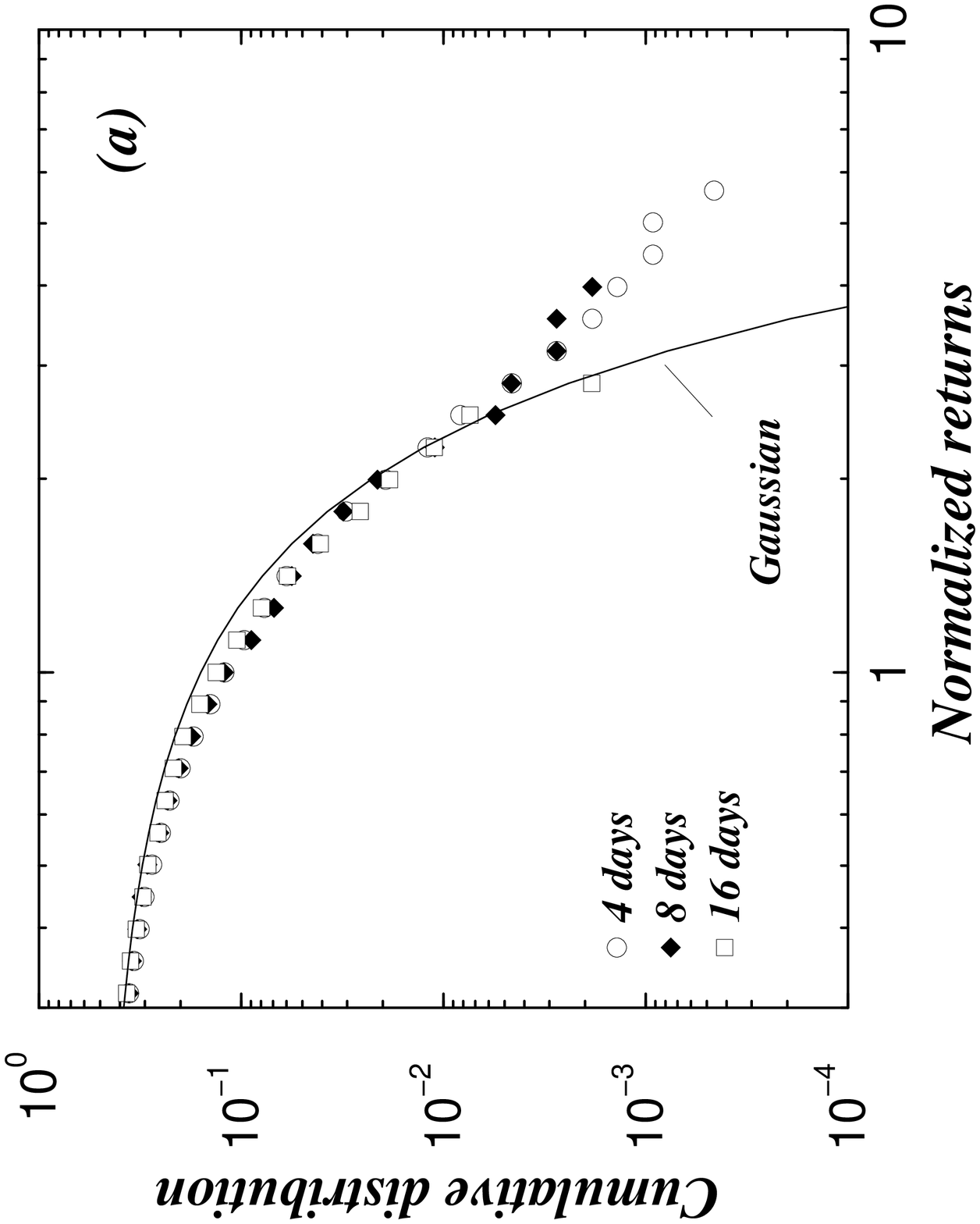}}}
}
\centerline{
\epsfysize=0.8\columnwidth{\rotate[r]{\epsfbox{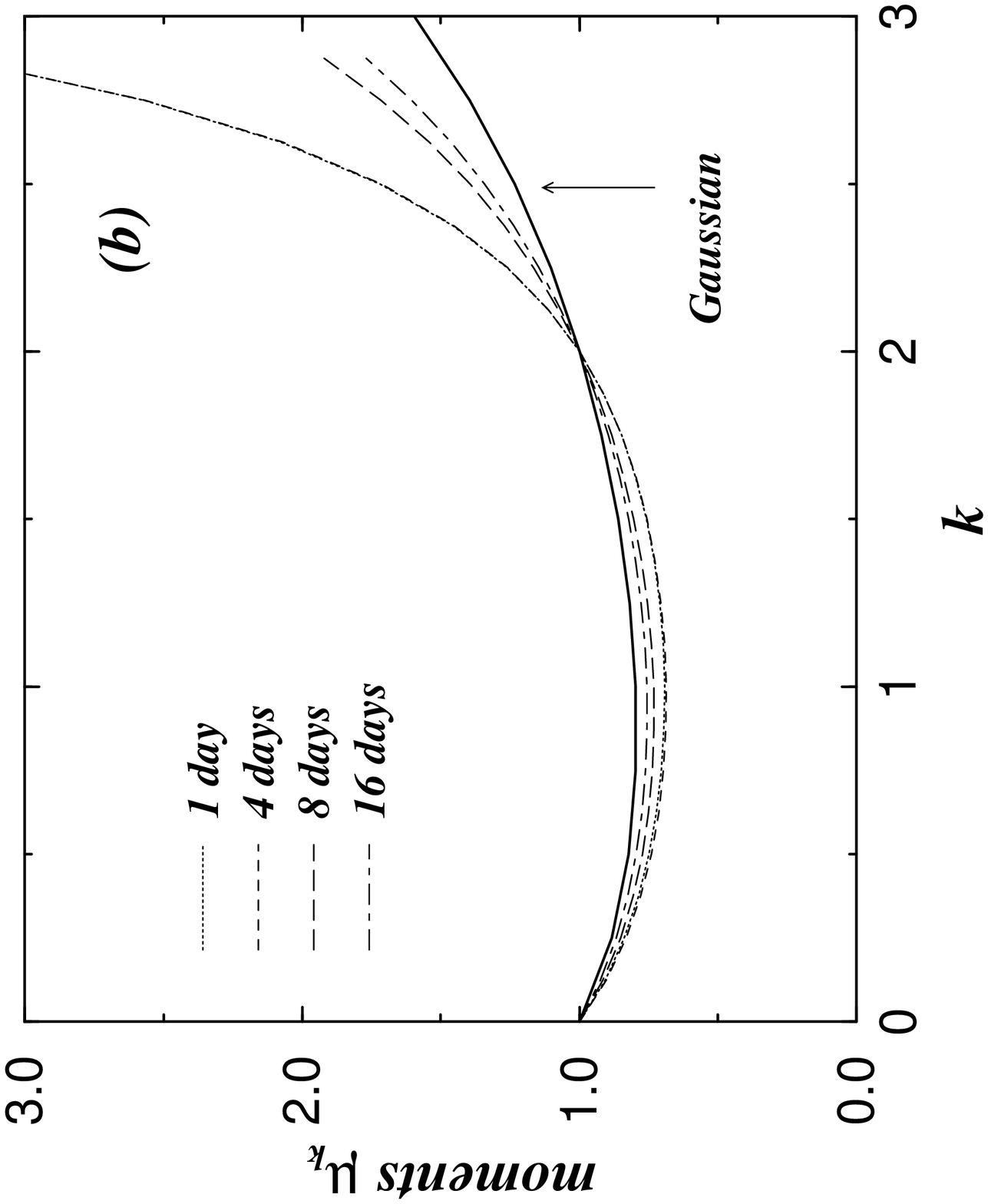}}} 
}
\caption{{\it(a)} Cumulative distribution for the positive tail of
S\&P 500 returns for time scales $\Delta t = 4, 8$ and 16~days. The
bold curve shows the cumulative distribution of a Gaussian with zero
mean and unit variance. {\it(b)} The moments for time scales $\Delta t
= 8$ and 16~days are consistent with a slow convergence to Gaussian
behavior. Note that the curves for $\Delta t = 1$ and 4 days are
indistinguishable. }
\label{sp_scl_brk}
\end{figure}

Figure~\ref{sp_scl_brk}(a) shows the distribution of the S\&P 500
returns for $\Delta t = 4, 8$ and 16~days. The data are now consistent
with a slow convergence to Gaussian behavior.  This is also visible
for the moments [Fig.~\ref{sp_scl_brk}(b)].

\subsection{The NIKKEI and Hang-Seng indices}

\begin{figure}
\centerline{ 
\epsfysize=0.8\columnwidth{\rotate[r]{\epsfbox{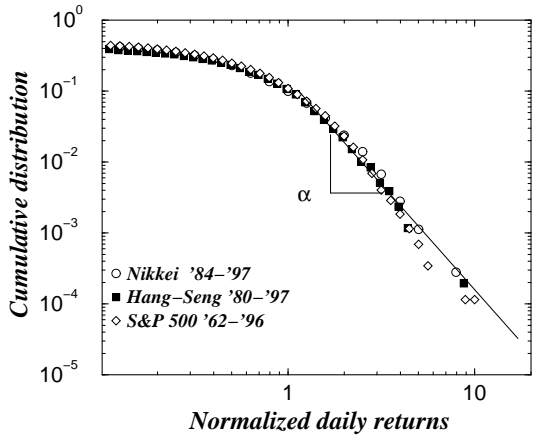}}} 
}
\caption{ Comparison of the cumulative distributions for the positive
tails of the normalized returns for the daily records of the NIKKEI
index from 1984-97, the daily records of the Hang-Seng index from
1980-97 and the daily records of the S\&P500 index. The apparent power
law behavior in the tails is characterized by the exponents $\alpha =
3.05 \pm 0.16$ (NIKKEI) , $\alpha = 3.03 \pm 0.16$ (Hang-Seng) and
$\alpha = 3.34 \pm 0.12$ (S\&P500). The fits are performed in the
region $g \ge 1$.}
\label{universality}
\end{figure}

The S\&P 500 is but one of the many stock market indices. Hence, we
investigate if the above results regarding the power-law asymptotic
behavior of the distribution of returns hold for other market indices
as well. Figure~\ref{universality} compares the distributions of daily
returns for the NIKKEI index of the Tokyo stock exchange and the
Hang-Seng index of the Hong Kong stock exchange with that of the S\&P
500. The distributions have similar functional forms, suggesting the
possibility of ``universal'' behavior of these distributions. In
addition, the estimates of $\alpha$ from regression fits,
\begin{equation}
\alpha = \cases{3.05 \pm 0.16  & (NIKKEI) \cr
3.03 \pm 0.16 & (Hang-Seng)},
\end{equation}
are in good agreement for the three cases.

\section{Dependence of average volatility on time scale}
 
The behavior of the time-averaged volatility $v(\Delta t)$ as a
function of the time scale $\Delta t$ is shown in
Fig.~\ref{acorsp}(c).  We find a power-law dependence,
\begin{equation}
v(\Delta t) \propto (\Delta t)^{\delta} .
\label{widthdelta} 
\end{equation}
We estimate $\delta \approx 0.7$ for time scales $\Delta t < 20\,$
min.  This value is larger than $1/2$ due to the exponentially-damped
time correlations, which are significant up to approximately
20~min. Beyond 20~min, $\delta \approx 0.5$, indicating the absence of
correlations in the returns, in agreement with
Fig.~\ref{acorsp}(a). The time-averaged volatility is also consistent
with essentially uncorrelated behavior for the daily and monthly
returns.

\section{Volatility Correlations and Time Scaling}

\begin{figure}
\narrowtext 
\centerline{
\epsfysize=0.8\columnwidth{\rotate[r]{\epsfbox{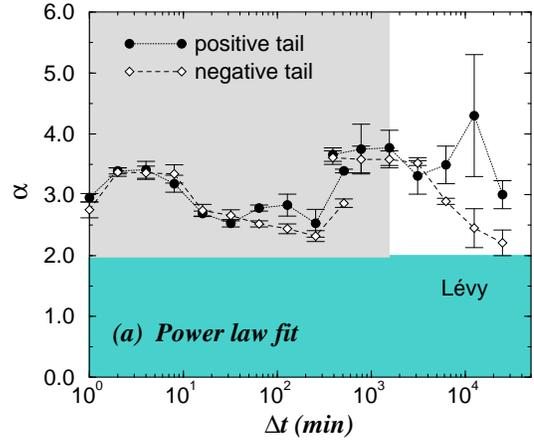}}}
}
\centerline{
\epsfysize=0.8\columnwidth{\rotate[r]{\epsfbox{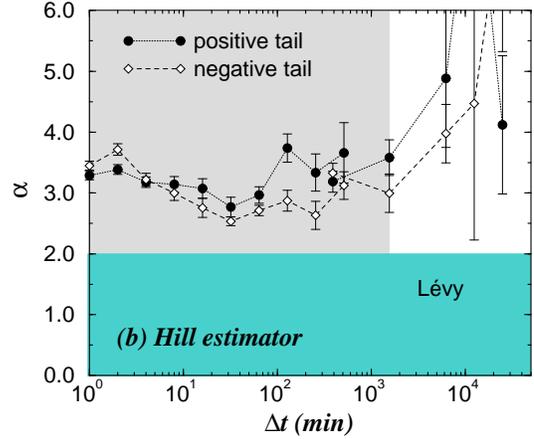}}}
}
\caption{ The values of the exponent $\alpha$ characterizing the
asymptotic power-law behavior of the distribution of returns as a
function of the time scale $\Delta t$ obtained using (a) a power-law
fit, and (b) the Hill estimator. The values of $\alpha$ for $\Delta t
< $1~day are calculated from database (i) which contains 13 years of
1~min records, while for $\Delta t \geq $1~day they are calculated
from database (ii), which has 35 years of daily records.  The unshaded
region, corresponding to time scales larger than $(\Delta t)_{\times}
\approx 4$~days (1560 min), indicates the range of time scales where
we find results consistent with slow convergence to Gaussian behavior
(see the text and the preceding figures).}
\label{alpha.fig}
\end{figure}

We have presented evidence that the distributions of returns retain
the same functional form for a range of time scales[see
Fig.~\ref{alpha.fig} and Table~\ref{alpha.exp}]. Here, we investigate
possible causes of this scaling behavior. Previous explanations of
scaling relied on L\'evy stable\cite{Mandelbrot63} and
exponentially-truncated L\'evy processes\cite{ms,Mantegna95}. However,
the empirical data that we analyze are not consistent with either of
these two processes.

\subsection{Rate of convergence}

Here, we compare the rate of convergence of the probability of the
returns to that of a computer-generated time series which has the same
distribution but is statistically independent by construction.  This
way, we will be able to study the convergence to Gaussian behavior of
independent random variables distributed as a power-law, with an
exponent $\alpha \approx 3$.

\begin{figure}
\centerline{ 
\epsfysize=0.8\columnwidth{\rotate[r]{\epsfbox{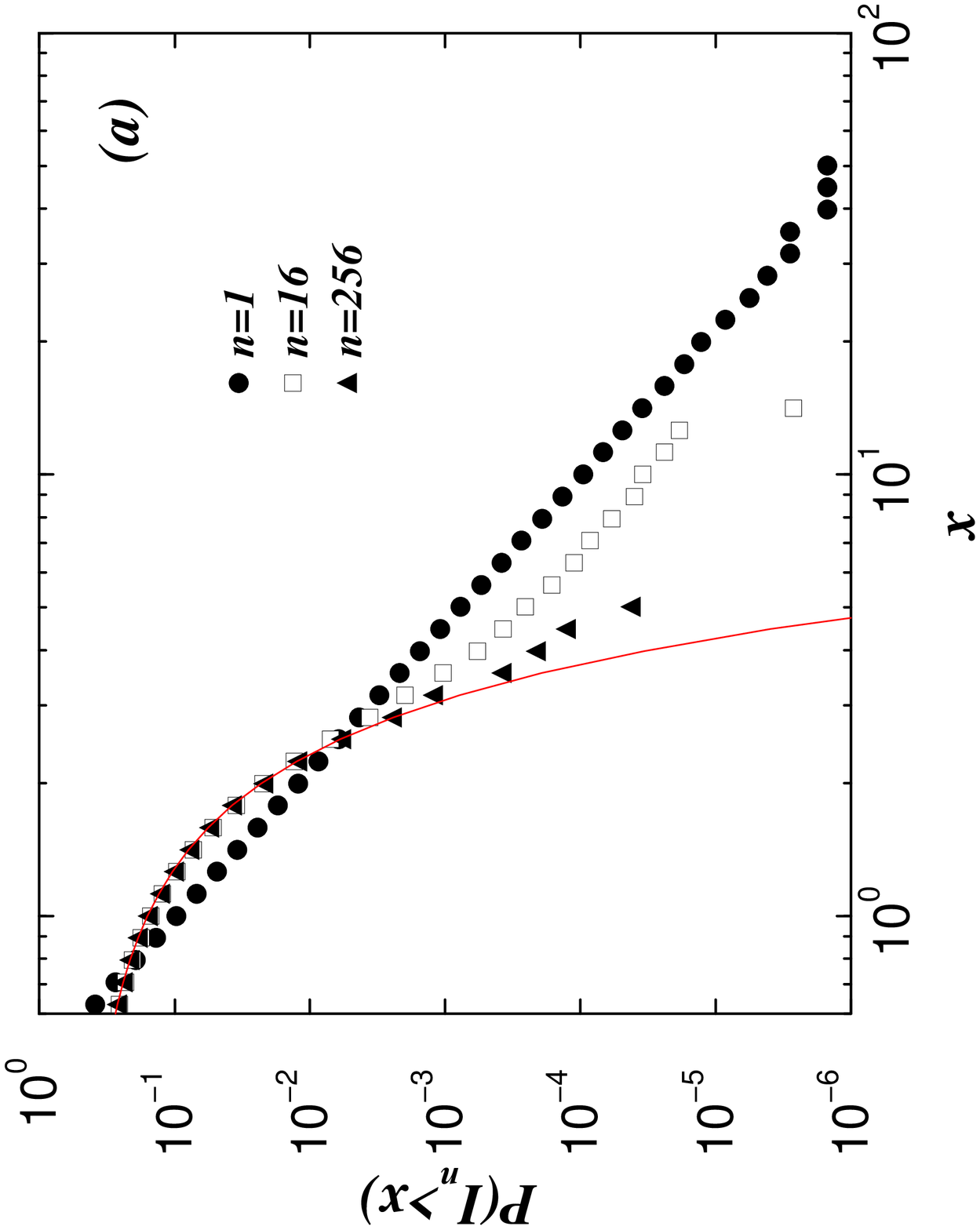}}}
}
\centerline{ 
\epsfysize=0.8\columnwidth{\rotate[r]{\epsfbox{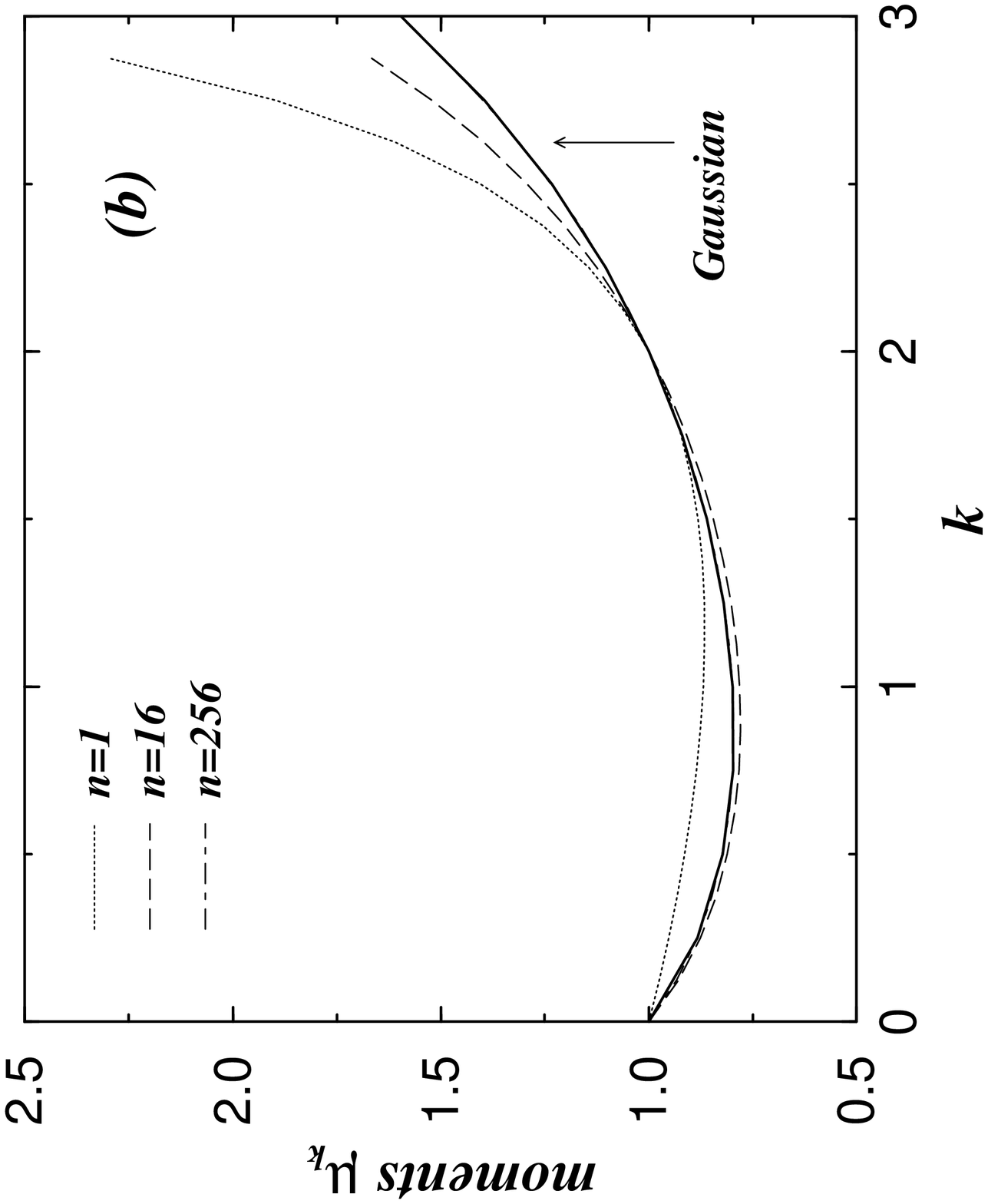}}}
}
\caption{ Convergence of distribution for independent variables.  We
first generate a time series $X_k$ distributed as $P(X\geq x) \sim
1/x^3$. We then generate the variables $I_n\equiv \sum_{i=1}^n X_k$
for $n=1,16$ and 256. {\it (a)} Cumulative distributions of
$I_n$. Note that the curve for $n=256$ is indistinguishable from the
Gaussian curve revealing convergence to Gaussian behavior.  {\it (b)}
The moments for $n=1,16$ and 256. These results can be compared with
Fig.~\protect\ref{sp_scl_brk}. Note that for the S\&P 500 even for
time scales $\Delta t = 16\,$days (corresponding to $n=208$) we still
do not observe a good degree of convergence. }
\label{sim_pow4}
\end{figure}

First, we generate a time series $X\equiv X_k\,,
k=1,\dots,40\times10^6 $ distributed as $P(X>x)\sim1/x^3$. We next
calculate the new random variables $I_n\equiv \sum_{i=1}^n X_k$, and
compute the cumulative distributions of $I_n$ for increasing values of
$n$.  These distributions show faster convergence with increasing $n$
than the distributions of returns [Fig.~\ref{sim_pow4}(a)]. This
convergence is also visible in the
moments. Figures~\ref{sim_pow4}(a,b) show that for $n=256$, both the
moments and the cumulative distribution show Gaussian behavior. In
contrast, for the distribution of returns, we observe significantly
slower convergence to Gaussian behavior: In the case of the S\&P 500
index, one observes a possible onset of convergence for $\Delta t
\approx$~4~days (1560~mins), starting from 1$\,$min returns.

These results confirm the existence of time dependencies in the
returns\cite{Dacarogna93,Ding93,Fama70,volatility,Granger96,Yanhui97,
Yanhui98,pierre,Cont97}.  Next, we show that the scaling behavior
observed for the S\&P 500 index no longer holds when we destroy the
dependencies between the returns at different times.

\subsection{Randomizing the time series of returns}

We start with the 1$\,$min returns and then destroy all the time
dependencies that might be present by shuffling the time series of
$G_{\Delta t=1}(t)$, thereby creating a new time series
$G_{1}^{sh}(t)$ which contains {\it statistically-independent\/}
returns. By adding up $n$ consecutive returns of the shuffled series
$G_{1}^{sh}(t)$, we construct the $n$$\,$min returns $G_{n}^{sh}(t)$.

\begin{figure}
\centerline{
\epsfysize=0.8\columnwidth{\rotate[r]{\epsfbox{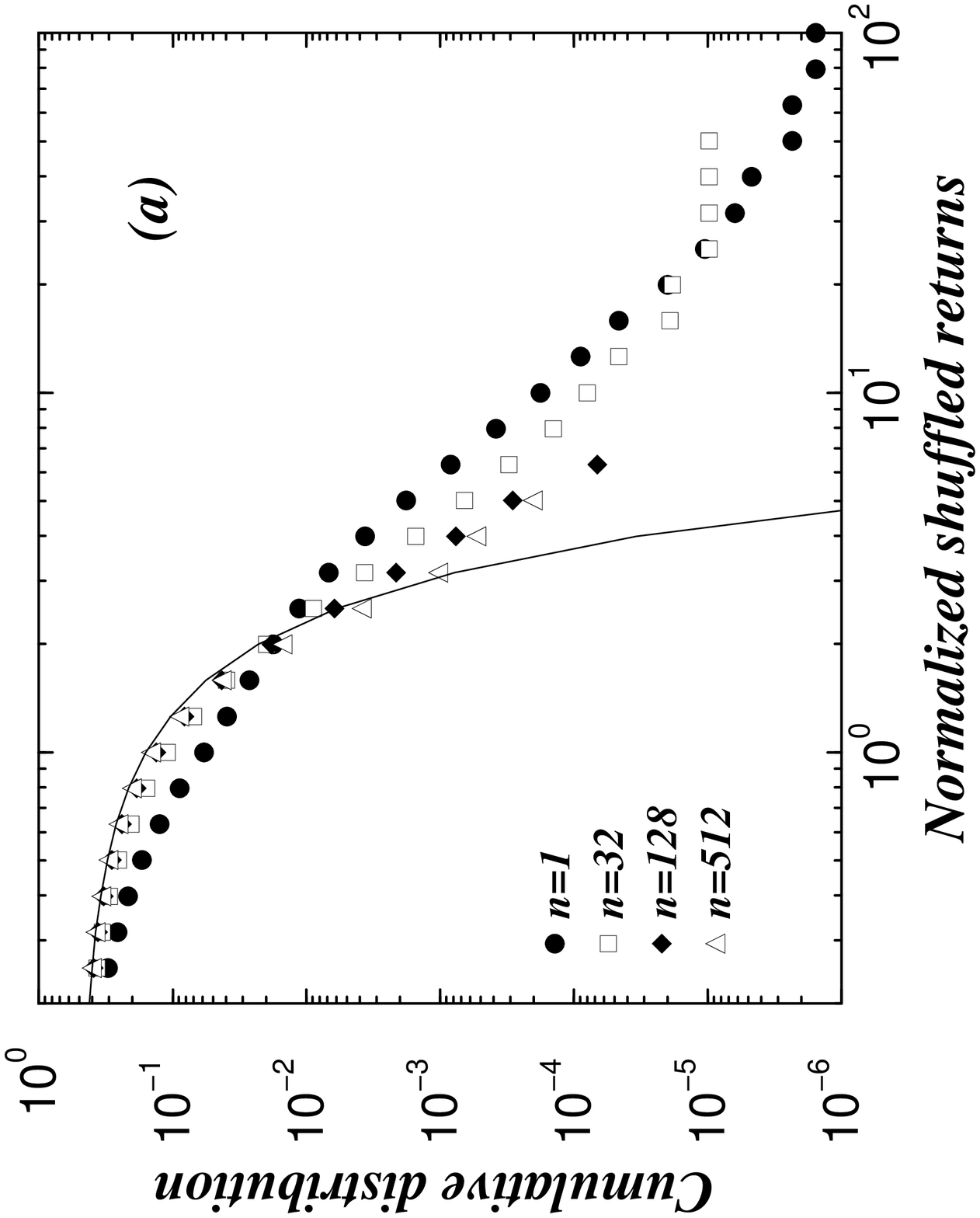}}}
}
\centerline{
\epsfysize=0.8\columnwidth{\rotate[r]{\epsfbox{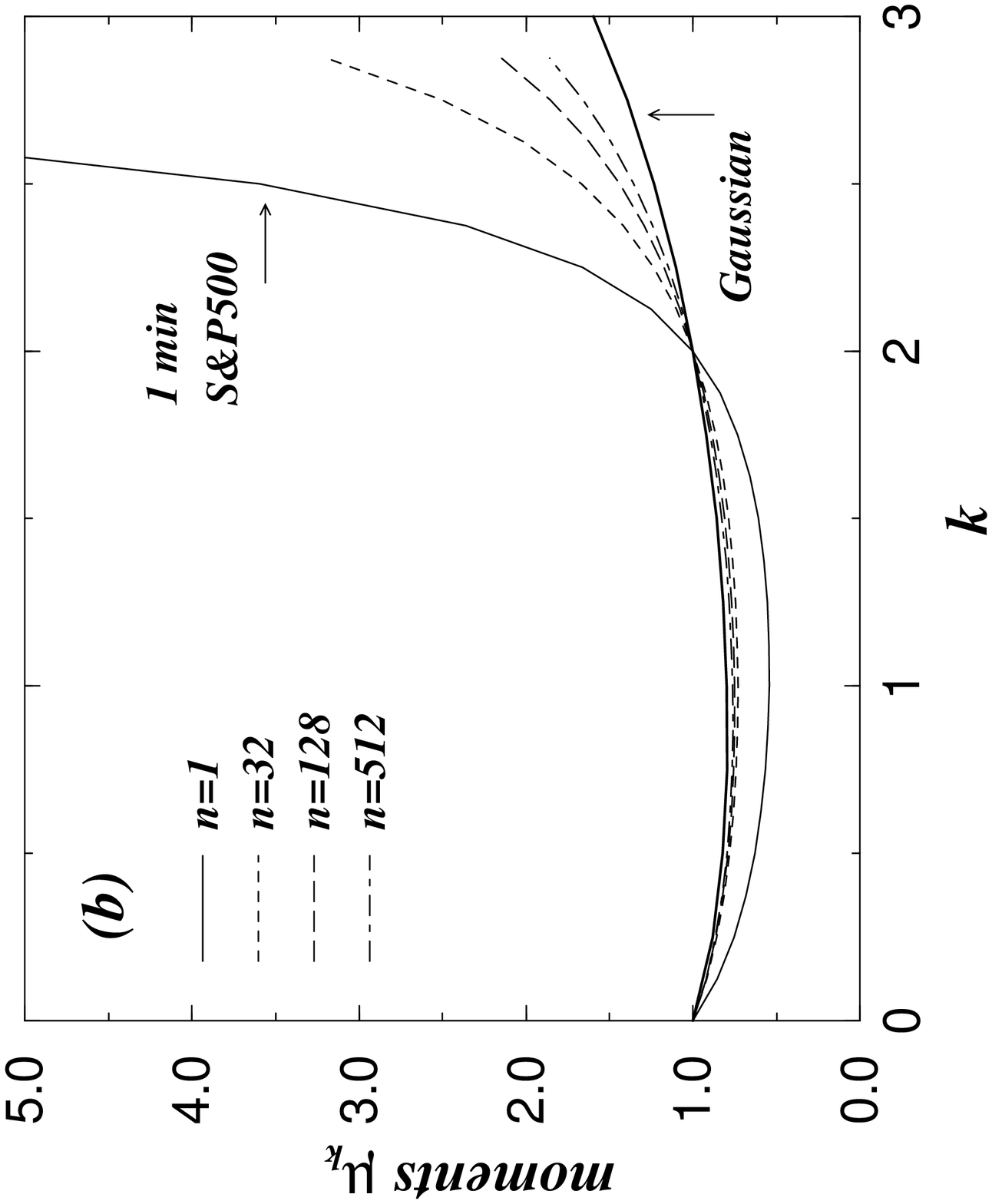}}}
}
\caption{ We randomize the time series of returns for the S\&P 500 for
$\Delta t$=1$\,$min and create a time series with the same
distribution but with independent random variables. We then sum up $n$
consecutive shuffled returns to create a shuffled $n$$\,$min
return. {\it(a)} Cumulative distributions of the positive tails of the
shuffled returns are shown for increasing $n$. We find slow
convergence to Gaussian behavior on increasing $n$. {\it (b)} The slow
convergence to a Gaussian behavior is shown by the moments. The
results in (b) can be compared with Fig.~\protect\ref{sim_pow4}(b) if
we note that $n=512$ corresponds to $\Delta t \approx 1.5\,$days. The
data are normalized to have the same second moment.}
\label{sp_shuffle}
\end{figure}

Figure~\ref{sp_shuffle}(a) shows the cumulative distribution of
$G_n^{sh}(t)$ for increasing values of $n$. We find a progressive
convergence to Gaussian behavior with increasing $n$. This convergence
to Gaussian behavior is also clear in the moments of $G_n^{sh}(t)$,
which rapidly approach the Gaussian values with increasing $n$
[Fig.~\ref{sp_shuffle}(b)].  This rapid convergence confirms that the
time dependencies cause the observed scaling behavior.

\section{Discussion}

We have presented a detailed analysis of the distribution of returns
for market indices, for time intervals $\Delta t$ ranging over roughly
4 orders of magnitude, from 1$\,$min up to 1 month
($\approx$~16,000~min).  We find that the distribution of returns is
consistent with a power-law asymptotic behavior, characterized by an
exponent $\alpha \approx 3$, well outside the stable L\'evy regime $0
< \alpha < 2$. For time scales $\Delta t \gg (\Delta t)_{\times}$,
where $(\Delta t)_{\times} \approx 4$~days, our results are consistent
with slow convergence to Gaussian behavior.

We have also demonstrated that the scaling behavior does not hold if
we destroy all the time dependencies by shuffling.  The breakdown of
the scaling behavior of the distribution of returns upon shuffling the
time series suggests that the long-range volatility correlations,
which persist up to several
months\cite{Dacarogna93,Ding93,Fama70,volatility,Granger96,Yanhui97,
Yanhui98,pierre,Cont97,Pasquini,Beran}, may be one possible reason for
the observed scaling behavior.

Recent studies~\cite{Yanhui98} show that the distribution of
volatility is consistent with an asymptotic power-law behavior with
exponent 3, just as observed for the distribution of returns. This
finding suggests that the process of returns may be written as
\begin{equation}
g(t) = \epsilon(t) \, v(t)\,,
\label{garch}
\end{equation}
where $g(t)$ denotes the return at time $t$, $v(t)$ denotes the
volatility, and $\epsilon(t)$ is an {\it i.i.d.} random variable {\it
independent\/} of $v(t)$. Since the asymptotic behavior of the
distributions of $v(t)$ and $g(t)$ is consistent with power-law
behavior, $\epsilon(t)$ should have an asymptotic behavior with faster
decay than either $g(t)$ or $v(t)$.  In fact, Eq.~(\ref{garch}) is
central to all the ARCH models~\cite{Engle82}, with $\epsilon(t)$
assumed to be Gaussian distributed.

Different ARCH processes assume different recursion relations for
$v(t)$. In the standard ARCH model, $v(t) = \alpha + \beta \,
g^2(t-1)$, leading to a power law distribution of returns with
exponent depending on the parameters $\alpha$ and $\beta$. However,
the standard ARCH process predicts a volatility correlation that
decays exponentially, since $v(t)$ depends only on the previous event,
and cannot account for the observed long-range persistence in
$v(t)$. To try to remedy this, one can require $v(t)$ to depend not
only on the previous value of $g(t)$ but on a finite number of past
events. This generalization is called the GARCH model.  Dependence of
$v(t)$ on the finite past leads not to a power-law decay (as is
observed empirically), but to volatility correlations that decay
exponentially ---with larger decay times as the number of events
``remembered'' is increased.

In order to explain the long range persistence of the autocorrelation
function of the volatility, one must assume that $v(t)$ depends on all
the past rather than a finite number of past
events~\cite{Ballie}. Such a description would be consistent with the
empirical finding of long-range correlations in the volatility, and
the observation that the distributions of $g(t)$ and $v(t)$ have
similar asymptotic forms. If the process of returns were governed by
the volatility, as in Eq.~(\ref{garch}), then the volatility would
seem to be the more fundamental process. In fact, it is possible that
the volatility is a measure of the amount of information arriving into
the market, and that the statistical properties of the returns may be
``driven'' by this information.

\section{Acknowledgments}

We thank J.-P. Bouchaud, M. Barth\'elemy, S.~V. Buldyrev, P. Cizeau,
X. Gabaix, I. Grosse, S. Havlin, K. Illinski, P.~Ch.~Ivanov, C.~King,
C.-K. Peng, B.~Rosenow, D. Sornette, D. Stauffer, S. Solomon, J.~Voit
and especially R.~N.~Mantegna for stimulating discussions and helpful
suggestions.  The authors also thank Bob Tompolski for his help
throughout this work. MM thanks DFG and LANA thanks FCT/Portugal for
financial support. The Center for Polymer Studies is supported by the
NSF.

\appendix

\section{L\'evy Stable Distributions}

L\'evy stable distributions arise from the generalization of the
central limit theorem to a wider class of distributions.  Consider the
partial sum $P_n\equiv\sum_{i=1}^nx_i$ of independent identically
distributed ({\it i.i.d.})  random variables $x_i$.  If the $x_i$'s
have finite second moment, the central limit theorem holds and $P_n$
is distributed as a Gaussian in the limit $n \rightarrow \infty$.

If the random variables $x_i$ are characterized by a distribution having
asymptotic power-law behavior
\begin{equation}
P(x)\sim x^{-(1+\alpha)}\,,
\end{equation}
where $\alpha < 2$, then $P_n$ will converge to a L\'evy stable
stochastic process of index $\alpha$ in the limit $n
\rightarrow \infty$.

Except for special cases, such as the Cauchy distribution, L\'evy
stable distributions cannot be expressed in closed form. They are
often expressed in terms of their Fourier transforms or characteristic
functions, which we denote $\varphi(q)$, where $q$ denotes the Fourier
transformed variable. The general form of a characteristic function of
a L\'evy stable distribution is
\begin{equation}
\ln \varphi(q) = \cases{ 
i\mu q-\gamma|q|^\alpha\left[1+i\beta{q\over|q|}tg\left(
{\pi\over 2}\alpha\right)\right] & $[\alpha\neq 1]$ \cr
i\mu q-\gamma|q|\left[1+i\beta{q\over |q|}{2\over\pi}\ln|q|\right] &
$[\alpha=1]$},
\end{equation}
where $0<\alpha\leq 2$, $\gamma$ is a positive number, $\mu$ is the
mean, and $\beta$ is an asymmetry parameter. For symmetric L\'evy
distributions ($\beta=0$), one has the functional form
\begin{equation}
P(x) = {1 \over 2\pi} \int_{-\infty}^\infty \, exp(-\gamma \vert q
\vert^{\alpha})\, e^{-iqx}\, dq \,.
\end{equation}
For $\alpha=1$, one obtains the Cauchy distribution and for the
limiting case $\alpha=2$, one obtains the Gaussian distribution.

By construction, L\'evy distributions are stable, that is, the sum of
two independent random variables $x_1$ and $x_2$, characterized by the
same L\'evy distribution of index $\alpha$, is itself characterized by
a L\'evy distribution of the same index. The functional form of the
distribution is maintained, if we sum up independent, identically
distributed L\'evy stable random variables.

For L\'evy distributions, the asymptotic behavior of $P(x)$ for $x\gg
1$ is a power-law,
\begin{equation}
P(x)\sim x^{-(1+\alpha)}\,.
\end{equation}
Hence, the second moment diverges. Specifically, $E\{|x|^n\}$ diverges
for $n\geq\alpha$ when $\alpha<2$.  In particular, all L\'evy stable
processes with $\alpha<2$ have {\it infinite\/} variance. Thus,
non-Gaussian stable stochastic processes do not have a characteristic
scale.  Although well-defined mathematically, these distributions are
difficult to use and raise fundamental problems when applied to real
systems where the second moment is often related to the properties of
the system.  In finance, an infinite variance would make risk
estimation and derivative pricing impossible.

\section{The Hill estimator (``local slopes'')}

A common problem when studying a distribution that decays as a power
law is how to obtain an accurate estimate of the exponent
characterizing the asymptotic behavior.  Here, we review the methods
of Hill\cite{Hill75}.  The basic idea is to calculate the {\it
inverse} of the local logarithmic slope $\zeta$ of the cumulative
distribution $P(g>x)$,
\begin{equation}
\zeta\,\equiv\, -\left( {d\log P \over d\log x} \right)^{-1} \,.
\label{def_gamma}
\end{equation}
We then estimate the inverse asymptotic slope $1/\alpha$ by
extrapolating $\zeta$ as $1/x \to 0$.  We start with the normalized
returns $g$ and proceed in the following steps:

{\bf Step I:} We sort the normalized returns $g$ in descending
order. The sorted returns are denoted $g_k$, $k=1, \dots, N$,
where $g_k > g_{k+1}$ and $N$ is the total number of events.

\begin{figure}
\centerline{
\epsfysize=0.8\columnwidth{\rotate[r]{\epsfbox{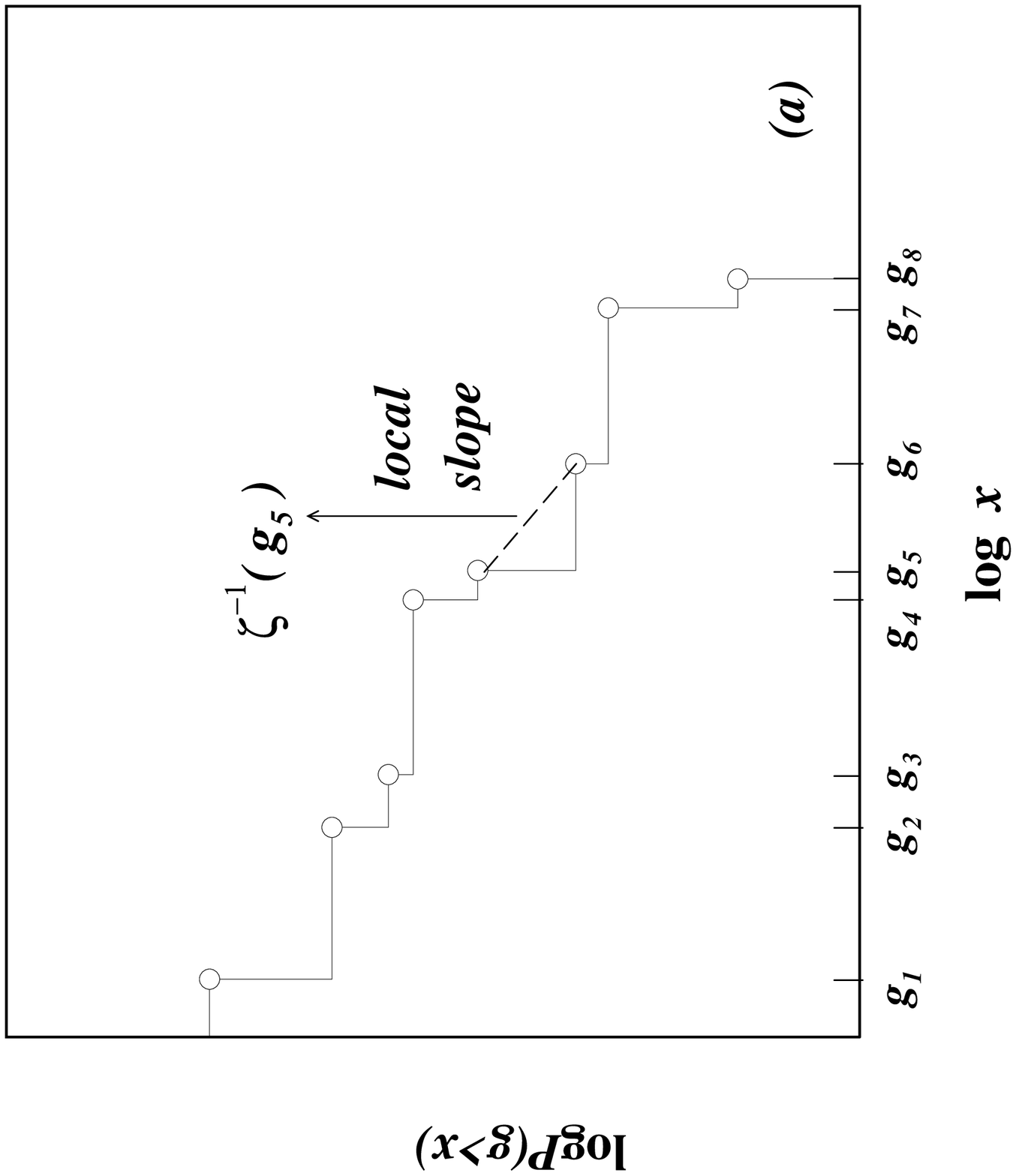}}}
}
\centerline{
\epsfysize=0.8\columnwidth{\rotate[r]{\epsfbox{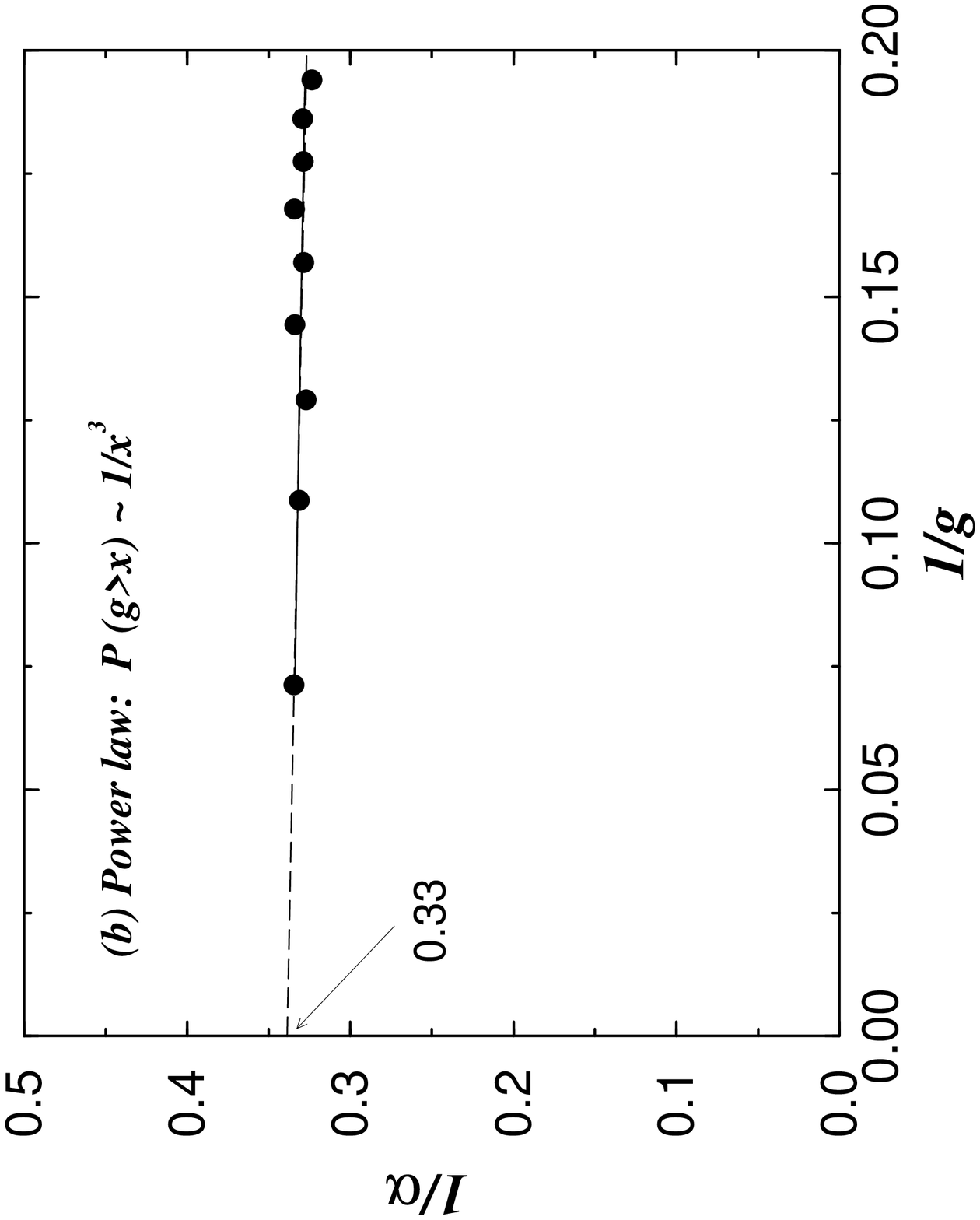}}}
}
\centerline{
\epsfysize=0.8\columnwidth{\rotate[r]{\epsfbox{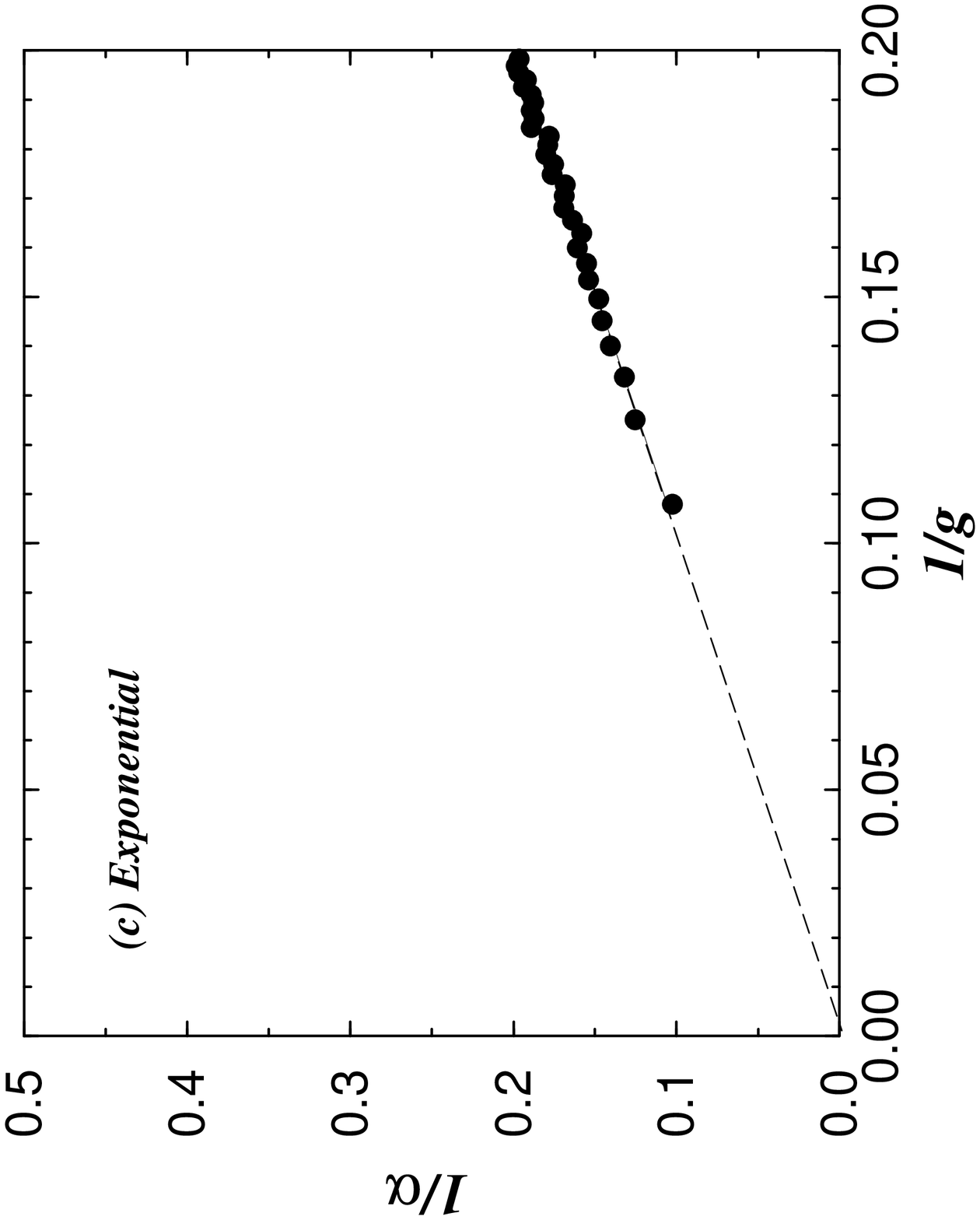}}}
}
\caption{{\it(a)} Schematic representation of the evaluation of the
local slope from the cumulative distribution.  First, the normalized
returns $g$ are sorted in descending order, $g_k > g_{k+1}$. The
dotted line indicates the local slope.  {\it (b)} Hill estimator for a
sequence of {\it i.i.d.\/} random variables with asymptotic behavior:
$P(g>x)=(1+x)^{-3}$. {\it(c)} Hill estimator for a sequence of {\it
i.i.d.\/} random variables with asymptotic behavior:
$P(g>x)=\exp(-x)$. Note that the asymptotic estimates, $1/\alpha =
0.33$ and $1/\alpha = 0$, recover for both cases the correct values of
$\alpha$, $\alpha=3$ and $\alpha=\infty$, respectively.}
\label{hill_explain}
\end{figure}

{\bf Step II:} The cumulative distribution is then expressed in terms
of the sorted returns as
\begin{equation}
P(g > g_k)={k \over N}\,.
\label{defP}
\end{equation}
Figure~\ref{hill_explain} is a schematic of the cumulative distribution
thus obtained. The {\it inverse} local slopes $\zeta(g)$ can be
written as
\begin{equation}
\zeta(g_k)=-{\log(g_{k+1}/g_k)\over
\log(P(g_{k+1})/P(g_k))} \,.
\label{def_gamm}
\end{equation}
Using Eq.~(\ref{defP}), the above expression can be well approximated
for large $k$ as
\begin{equation}
\zeta(g_k) \simeq k(\log(g_{k+1})-\log(g_k))\,,
\label{def_gamm1}
\end{equation}
yielding estimates of the local inverse slopes.

\vspace{0.5cm}

{\bf Step III:} We obtain the inverse local slopes through
Eq.~(\ref{def_gamm1}).  We can then compute an average of the inverse
slopes over $m$ points,
\begin{equation}
\langle \zeta \rangle \equiv {1\over m} \sum_{k=1}^m \zeta(g_k),
\label{def_avgamma}
\end{equation}
where the choice of the averaging window length $m$ varies depending
on the number of events $N$ available.

\begin{figure} 
\centerline{ 
\epsfysize=0.8\columnwidth{\rotate[r]{\epsfbox{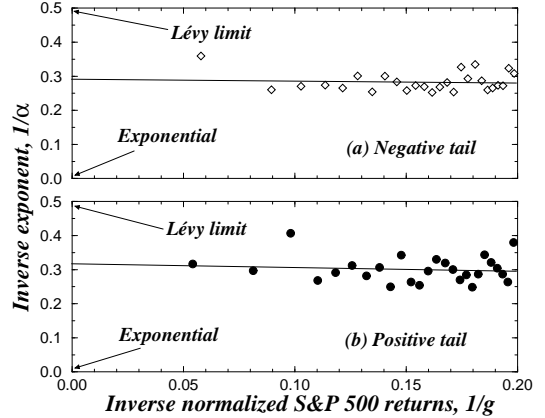}}} 
} 
\caption{ Inverse local slopes of the cumulative distributions of
normalized returns for $\Delta t = 1 min$ for the {\it (a)} positive
and {\it (b)} negative tails. Each point is an average over 100
different inverse local slopes. Extrapolation of the regression lines
provides estimates for the asymptotic slopes $\alpha = 3.45 \pm 0.07$
(positive tail), and $\alpha = 3.29 \pm 0.07$ (negative tail).}
\label{sp500_hill} 
\end{figure}

{\bf Step IV:} We plot the locally averaged inverse slopes $\langle
\zeta \rangle$ obtained in Step III as a function of the inverse
normalized returns $1/g$ [see, e.g., Fig.~\ref{sp500_hill}]. We can
then define two methods of estimating $\alpha$.  In the first method,
we extrapolate $\zeta$ as a function of $1/g$ to $0$, similarly to the
method of successive slopes\cite{Stanley67}; this procedure yields the
inverse asymptotic slope $1/\alpha$.  In the second method, we average
over all events for $1/g$ smaller than a given
threshold~\cite{Hill75}, with the average yielding the inverse slope
$1/\alpha$.
 
To test the Hill estimator, we analyze two surrogate data sets with
known asymptotic behavior: (a) an independent random variable with
$P(g>x)=(1+x)^{-3}$, and (b) an independent random variable with
$P(g>x)=\exp(-x)$. As shown in Figs.~\ref{hill_explain}(b,c), the
method yields the correct results $\alpha=3$ and $\alpha=\infty$,
respectively.


\begin{table}[hbt]
\narrowtext
\caption{ The values of the exponent $\alpha$, for different time
scales $\Delta t$, for the S\&P 500 index: (a) power-law regression
fit to the cumulative distribution, and (b) Hill estimator.  The
daggered values are computed using database (ii), which contains
daily-sampled records, while the values without the dagger are
computed using database (i), which contains records with a 1$\,$min
sampling.  Note that we use the conversion
1~day=390~min.~\protect\cite{daymin}. \protect\vspace*{1cm} }

\begin{tabular}{cllll}
\multicolumn{1}{c}{$\Delta t~(min)$}    
        &\multicolumn{2}{c}{Power law fit} 
        & \multicolumn{2}{c}{Hill estimator}\\ \cline{2-5} 
                &Positive       &Negative       &Positive       &Negative \\
\hline
1     &$2.95 \pm 0.07$ &$2.75 \pm 0.13$ &$3.29 \pm 0.07$ &$3.45 \pm 0.07$\\
2     &$3.39 \pm 0.05$ &$3.37 \pm 0.07$ &$3.38 \pm 0.08$ &$3.71 \pm 0.09$\\
4     &$3.41 \pm 0.14$ &$3.36 \pm 0.11$ &$3.18 \pm 0.09$ &$3.22 \pm 0.10$\\
8     &$3.18 \pm 0.14$ &$3.34 \pm 0.15$ &$3.14 \pm 0.13$ &$3.00 \pm 0.12$\\
16    &$2.69 \pm 0.04$ &$2.74 \pm 0.10$ &$3.07 \pm 0.26$ &$2.75 \pm 0.16$\\
32    &$2.53 \pm 0.06$ &$2.66 \pm 0.09$ &$2.77 \pm 0.16$ &$2.53 \pm 0.07$\\
64    &$2.78 \pm 0.05$ &$2.52 \pm 0.05$ &$2.97 \pm 0.14$ &$2.71 \pm 0.09$\\
128   &$2.83 \pm 0.18$ &$2.44 \pm 0.08$ &$3.74 \pm 0.23$ &$2.87 \pm 0.17$\\
256   &$2.53 \pm 0.23$ &$2.32 \pm 0.09$ &$3.33 \pm 0.30$ &$2.63 \pm 0.23$\\
390$^{\dagger}$   &$3.66 \pm 0.11$ &$3.61 \pm 0.11$ &$3.19 \pm 0.17$ &$3.33 \pm 0.16$\\
512   &$3.39 \pm 0.03$ &$2.86 \pm 0.07$ &$3.7  \pm 0.5$  &$3.12 \pm 0.23$\\
780$^{\dagger}$   &$3.75 \pm 0.41$ &$3.58 \pm 0.22$ &$3.06 \pm 0.26$ &$4.67 \pm 0.38$\\
1560$^{\dagger}$  &$3.77 \pm 0.29$ &$3.58 \pm 0.14$ &$3.58 \pm 0.29$ &$2.99 \pm 0.32$\\
3120$^{\dagger}$  &$3.31 \pm 0.30$ &$3.52 \pm 0.04$ &$4.9  \pm 0.6$  &$3.85 \pm 0.45$\\
6240$^{\dagger}$  &$3.49 \pm 0.31$ &$2.89 \pm 0.05$ &$4.9  \pm 1.1$  &$3.97 \pm 0.48$\\
12480$^{\dagger}$ &$4.3 \pm 1.0$ &$2.45 \pm 0.32$ &$8.7  \pm 2.0$  &$4.5  \pm 2.2$ \\
24960$^{\dagger}$ &$3.00 \pm 0.23$ &$2.21 \pm 0.21$ &$4.1  \pm 1.1$  &$7.7  \pm 2.4$ \\
\end{tabular}
\label{alpha.exp}
\end{table}


\end{multicols} 


\begin{references}
\bibitem{frc} Some previous studies\protect{\cite{Mantegna95}} were
conducted with the return defined as the difference in the price and
not the difference in the logarithm of the price. However, for events
that are not too large, the two quantities have the same statistics.
\bibitem{Nonstationarity} Non-stationarity in the time series of
returns is an issue which has been investigated extensively (see
\cite{Pagan96} for a summary).
\bibitem{sp_comm} The S\&P 500 index is the sum of the market
capitalizations (share price multiplied with the number of outstanding
shares) of 500 companies representing the major industries. Note that
these are not necessarily the largest, and that the composition of the
index changes slowly in time.
\bibitem{FXcorr} For foreign exchange markets the decay time is even
smaller\protect\cite{Dacarogna93}.
\bibitem{discrete} The discreteness in the returns is due to the fact
that stocks are priced in units of fractions of USD, usually 1/8,
1/16, or 1/32. The discreteness is not so evident in the S\&P 500
index as it is the sum of 500 stocks, each with different weights.
\end{references}
\end{document}